\documentclass[fleqn,usenatbib]{mnras}

\usepackage{newtxtext,newtxmath}

\usepackage[T1]{fontenc}
\usepackage{ae,aecompl}

\usepackage{graphicx}	
\usepackage{amsmath}	
\usepackage{amssymb}	
\usepackage{multirow}
\usepackage{multicol}
\usepackage[section]{placeins}
\usepackage{tabularx}
\usepackage{nameref}
\usepackage{longtable}
\usepackage[T1]{fontenc}
\usepackage{fix-cm}
\usepackage{siunitx}
\usepackage{caption}
\usepackage{float}
\restylefloat{figure}
\usepackage{dblfloatfix}    

\def\parcm{\hbox{$.\!^{\prime}$}}
\def\parcs{\hbox{$.\!\!^{\prime\prime}$}}
\def\surfb{{$\,$\text{mag}/$\Box^{\prime\prime}$}}

\usepackage{etoolbox}

\title[LBT-SONG: The Satellite Population of NGC 628]{The LBT Satellites of Nearby Galaxies Survey (LBT-SONG): The Satellite Population of NGC 628}

\author[A. Bianca Davis et al.]{
A. Bianca Davis$^{1,3}$\thanks{E-\text{M}ail: davis.4811@osu.edu},
Anna M. Nierenberg,$^{4,5}$,
Annika H. G. Peter$^{1,2,3}$,
\newauthor
Christopher T. Garling$^{2,3}$,
Johnny P. Greco$^{1,3,7}$,
Christopher S. Kochanek$^{2,3}$,
\newauthor
Dyas Utomo$^{2}$,
Kirsten Casey$^{1,3}$,
Richard W. Pogge$^{2,3}$,
Daniella Roberts$^{1,3}$,
\newauthor
David J. Sand$^{6}$,
Amy Sardone$^{2,3,7}$
\\
$^{1}$Department of Physics, The Ohio State University, 191 W. Woodruff Ave., Columbus OH 43210, USA\\
$^{2}$Department of Astronomy, The Ohio State University, 140 W. 18th Ave., Columbus OH 43210, USA\\
$^{3}$Center for Cosmology and Astroparticle Physics, The Ohio State University, 191 W. Woodruff Ave., Columbus OH 43210, USA\\
$^{4}$Jet Propulsion Laboratory, California Institute of Technology\\
$^{5}$NASA Postdoctoral Program Fellow\\
$^{6}$Steward Observatory, University of Arizona, 933 North Cherry Avenue, Rm. N204, Tucson, AZ 85721-0065, USA\\
$^{7}$NSF Astronomy and Astrophysics Postdoctoral Fellow
}

\date{Accepted XXX. Received YYY; in original form ZZZ}

\pubyear{2020}

\begin{document}
\setlength{\parskip}{0pt}
\label{firstpage}
\pagerange{\pageref{firstpage}--\pageref{lastpage}}
\maketitle

\begin{abstract}
We present the first satellite system of the Large Binocular Telescope Satellites Of Nearby Galaxies Survey (LBT-SONG), a survey to characterize the close satellite populations of Large Magellanic Cloud to Milky Way-mass, star-forming galaxies in the Local Volume. In this paper, we describe our unresolved diffuse satellite finding and completeness measurement methodology and apply this framework to NGC 628, an isolated galaxy with $\sim1/4$ the stellar mass of the Milky Way. We present two new dwarf satellite galaxy candidates: NGC 628 dwA, and dwB with $\text{M}_{\text{V}}$ = $-12.2$ and $-7.7$, respectively. NGC 628 dwA is a classical dwarf while NGC 628 dwB is a low-luminosity galaxy that appears to have been quenched after reionization. Completeness corrections indicate that the presence of these two satellites is consistent with CDM predictions. The satellite colors indicate that the galaxies are neither actively star-forming nor do they have the purely ancient stellar populations characteristic of ultrafaint dwarfs. Instead, and consistent with our previous work on the NGC 4214 system, they show signs of recent quenching, further indicating that environmental quenching can play a role in modifying satellite populations even for hosts smaller than the Milky Way. 

\end{abstract}




\section{Introduction}

The $\Lambda$ cold dark matter (CDM) cosmological model predicts the existence of a hierarchy of dark matter halos, in the centers of which galaxies form and reside (see \citealt{WechslerTinker2018} for a review). This model has had many successes on large scales. For example, dark-matter-only CDM N-body simulations produce a network of halo structures in remarkable statistical agreement with the spatial distribution and evolution of massive galaxies ($\text{M}_{\rm halo}\gtrsim 10^{12} \text{M}_{\odot}$; \citealt{schaye2015,Vogelsberger2014,Zu2015}). The global star formation histories (SFH) of large galaxies are well-described by models which use a simple abundance matching prescription to assign galaxies to halos after ranking both by mass \citep[]{Tasitsiomi_2004,Conroy_2006,Behroozi13,Moster13}. However, in the dwarf galaxy regime below this mass scale ($\text{M}\lesssim10^{10} \text{M}_{\odot}$), the number, masses and densities of galaxies predicted by the $\Lambda$CDM model are not in clear agreement with observations. This has led researchers to grapple with the `missing satellites' \citep[e.g.,][]{kauffmann93,Moore99,KlypinMissingSats}, `too big to fail' \citep{Boylan_Kolchin_2011} and `cusp-core' \citep[e.g.,][]{Flores_1994,Moore1994,NFW1997,Moore_1999} problems of $\Lambda$CDM on small scales \citep{Weinberg2015,Bullock_2017}.

One solution to these problems is to modify dark matter particle properties in a way that would change the number of small halos, their expected masses, and densities \citep{Hu2000,Vogelsberger2016,Hui2017,Lovell2017}. On the other hand, other probes of small-scale structure which do not rely on detecting luminous baryons in halos generally show good consistency with CDM. Strong lensing and Lyman-$\alpha$ forest measurements, as well as measurements of galaxy abundances in the local Universe, show halo mass functions match CDM predictions at mass scales of $10^{7-8} \text{M}_{\odot}$ \citep[e.g.,][]{weinberg1997,Dalal_2002,Strigari_2007,Tollerud2008,Behroozi_2010,Reddick_2013,Viel_2013,Baur_2016,Jethwa_2017,Irsic2017,Kim2018,Nadler_2019,gilman2019warm,nierenberg2019double}. While there is still a window open for some dark-matter solutions to small-scale structure problems (especially in the context of halo structure rather than abundance), non-DM physics is a more likely solution.

Most of the problems for low mass halos seem to arise because the physics of galaxy formation is not yet sufficiently well understood in the context of small dark-matter halos. Our uncertainties can be cast in terms of the mapping between galaxy stellar and halo masses, the $\text{M}_{\star}-\text{M}_{\rm halo}$ relation, which is not well-constrained by data below $M_{\text{halo}} = 10^{10} \text{M}_{\odot}$. In simulations there is significant variance in the mean relation and in the scatter for halo masses below $M_{\text{halo}} = 10^{10} \text{M}_{\odot}$ \citep{Munshi13,munshi2019,brooksZolotov2014,sawalaBentByBaryons2015,Simpson2015,Fitts2017,Wheeler2019}. Predictions of the luminous satellite populations of galaxies are subject to a high degree of uncertainty even for a well-understood subhalo mass function.

Various global, internal, and environmentally-dependent astrophysical processes may disproportionately influence the formation of stars in dwarf galaxies compared to their more massive counterparts over cosmic time, complicating the mapping between the luminous dwarf galaxies and their halos. Understanding these processes is important both for understanding the physics of galaxy formation, and determining how much room is left for novel dark-matter physics on dwarf galaxy scales. 

Globally, reionization could quench star formation on small scales, leaving halos below some mass threshold dark and completely devoid of luminous baryons \citep{Barkana1999,Gnedin2000,Bullock00,Benson02,Read2005,rodriguezwimberly2019}. We see evidence of this in a new class of `Ultrafaint' dwarfs (UFDs) with $\text{M}_{\star}\lesssim10^5\text{M}_{\odot}$, which have been discovered in the Local Group \citep[e.g.,][]{Zucker04,Willman05a,Belokurov07,Laevens2015,Kim2015,DrlicaWagner2015,Torrealba16a,Torrealba2018a,Torrealba2018b,Homma2018}. Star formation in these UFDs ceased much earlier than in the more massive `classical' dwarfs ($10^5 \text{M}_{\odot} \lesssim \text{M}_{\star} \lesssim 10^9 \text{M}_{\odot}$), consistent with quenching by reionization \citep{weisz2014b,brown2014,simon2019}.  

Internal baryonic processes may disproportionately alter dwarf galaxies relative to their more massive counterparts as they have shallower potential wells. Stellar winds and supernovae feedback could remove cold gas and quench star formation more efficiently, leading to a suppression of star formation compared to larger galaxies \citep{Tollerud2011,Simpson2015,emerick2016gas}. In addition, internal feedback mechanisms may also modify the density profiles of the dark matter halos \citep{pontzen2012,brooksZolotov2014,Read2019}, changing the mapping between galaxy kinematics and the inferred halo mass, and further altering the $\text{M}_{\star}-\text{M}_{\rm halo}$ relation. 

There are also indications that environmental effects play an important role in shaping the luminosity function and lives of classical dwarf galaxies, and that these effects depend on satellite galaxy properties. Environmental quenching mechanisms such as strangulation \citep{larson1980}, ram-pressure \citep{gunn1972} and tidal stripping can also affect dwarf galaxy star formation and quenching times as a function of host and satellite mass and orbit \citep{mayer2006,nichols2011,Wetzel:2016wro,digby2019}. With the exception of the Large and Small Magellanic clouds, all Milky Way satellites are quenched, while field dwarfs are star-forming \citep{geha2012}, suggesting that environment plays a crucial evolutionary role for satellites of Milky Way-mass hosts \citep[see also ][]{slater2014,wetzel2015,fillingham2016}. This behavior has been observed in other Mikly Way-mass and larger systems such as M31 \citep{McConnachieM31}, M81 \citep{chiboucas2013confirmation}, CenA \citep{crnojevic2019faint}, and M101 \citep{bennet2019m101}. The hot accretion-shocked gas halos around Milky Way-mass hosts were thought to be necessary to quench satellites through strangulation and ram pressure stripping. Lower mass galaxies, which are believed to not have hot coronae \citep{Correa_2017}, have recently been been shown to have quenched satellites \citep{garling2019ddo113}, indicating that intermediate and low mass hosts can also quench star formation in satellites.

Simple semi-empirical models for the $\text{M}_{\star}-\text{M}_{\rm halo}$ relation coupled with reionization suppression and standard CDM halo mass functions show good agreement with the Milky Way's satellite population  \citep{Koposov2008,Tollerud2008,Walsh2009,Hargis2014,Jethwa_2017,Kim2018,Newton2018,Nadler2019}. This appears to solve the `missing satellites problem' \citep{Klypin99,Moore99}, although there may still be a deficit of large stellar mass classical dwarfs compared to theoretical predictions \citep{Brooks2013,Dooley17b,Kim2018}. However, these models do not incorporate environmental effects, and there is some concern that these semi-empirical models are ``over-tuned" to the Milky Way \citep{geha2017saga,Kim2018}.

\begin{figure*}
 \includegraphics[width=\textwidth]{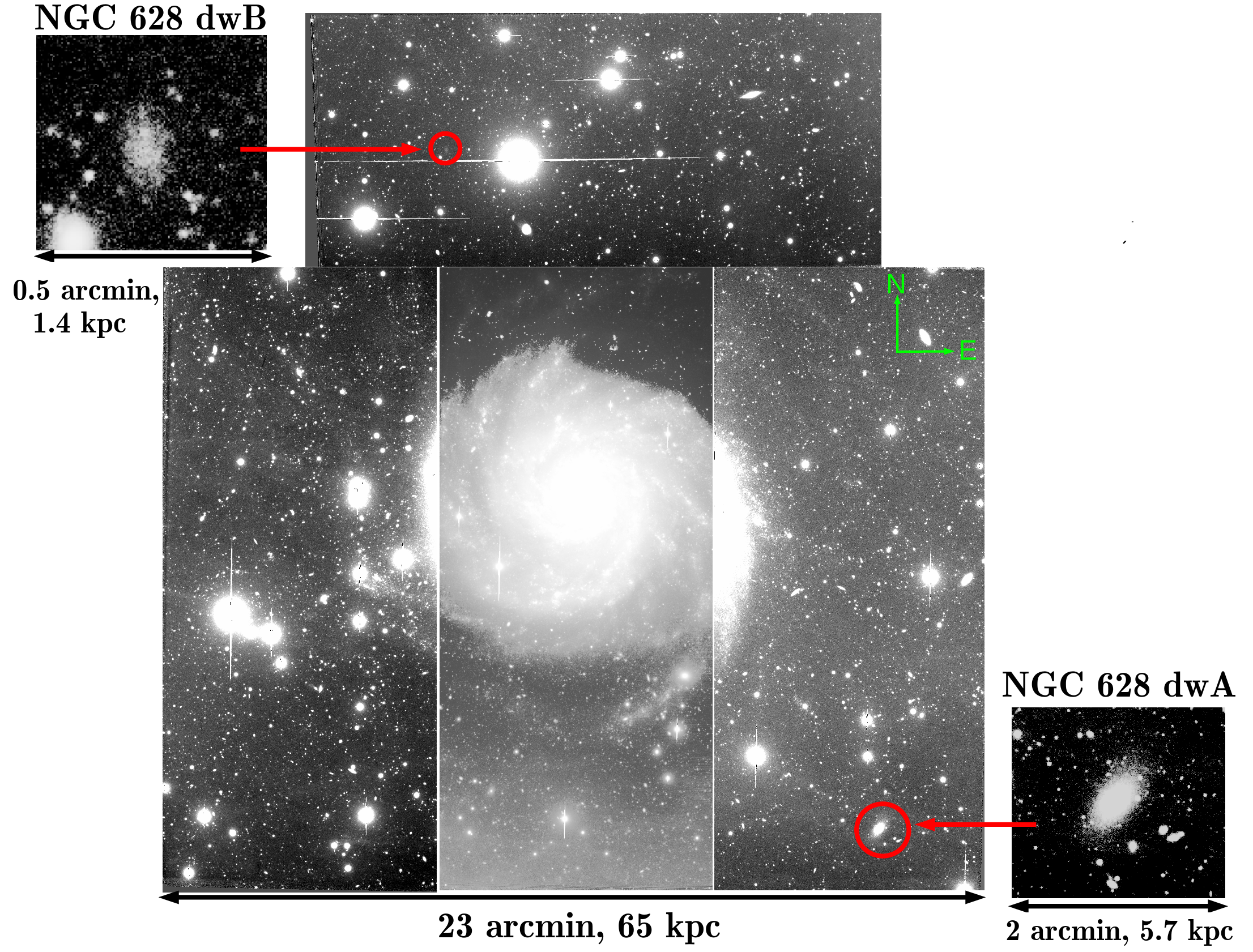}
 \caption{The stacked R-band LBC image of NGC 628. On the top and right-most panels, the two candidates NGC 629 dwA and dwB positions are circled, with zoomed in images shown adjacent.}
  \label{fig:Host}
\end{figure*}

Therefore, to disentangle the various baryonic processes, and to make better models of, the $\text{M}_{\star}-\text{M}_{\rm halo}$ relation as a function of environment, it is necessary to obtain good statistical samples of satellite systems in a variety of environments. Beyond the Local Group, large optical galaxy surveys such as the Sloan Digital Sky Survey \citep{york2000} and the Galaxy And Mass Assembly (GAMA) survey \citep{Gama} search for galaxies but are limited by magnitude and surface brightness limits to galaxies similar in luminosity to the Fornax dwarf galaxy ($\text{M}_{\star}\sim10^7 \text{M}_{\odot}$) and larger. While gas-rich dwarfs in the field are found in H\textsc{i} surveys \citep{papastergis2012,tollerud2014,bernstein2014alfalfa,cannon2015alfalfa,sand2015comprehensive}, this method cannot find reionization-fossil or quenched galaxies typical of satellites of the Milky Way due to the absence of gas. The number of small halos and their relationship to galaxies on these mass scales outside the Milky Way is therefore poorly known \citep{Loveday2015}. 

New optical surveys are targeting the satellite populations of hosts with $\text{M}_{\star}$ larger than $10^7 \text{M}_{\odot}$ to uncover the satellite galaxy luminosity functions below that scale. These surveys have mostly targeted denser cluster environments \citep{Grossauer15,van2015forty}, Milky-Way-sized hosts \citep{merritt2014discovery,geha2017saga,bennet2019m101}, and Large Magellanic Cloud-sized hosts \citep{Carlin16}. More recent work has targeted the satellites of nearby LMC to Milky Way-mass galaxies in different environments \citep{Carlsten_2020}.

This work presents the first satellite system of the Large Binocular Telescope Satellites Of Nearby Galaxies (LBT-SONG) program to find and characterize the dwarf satellite populations of intermediate mass host galaxies outside the Local Group. Our project will report candidates for each host as well as completeness calculations for dwarfs as low in mass as the Hercules dwarf spheroidal ($\text{M}_{\star} \sim10^4\text{M}_{\odot}$). Our hosts range in distance from 3 to 10 Mpc. We apply two different techniques of satellite detection based on the distance. Those within 5 Mpc are analyzed using a resolved stellar population analysis (Garling et al. in prep.) while those beyond 5 Mpc are searched for as faint ``diffuse sources" and use the approach detailed in this paper.

We begin our study of the ``diffuse dwarf" host sample with NGC 628, at a distance of 9.77 Mpc \citep{Mcquinn2017Distance}, where we identify two dwarf galaxy candidates and estimate the completeness of the search. In Section 2, we describe the survey data and our data reduction procedure. In Section 3, we describe our diffuse source detection pipeline, including our mock galaxy injection and recovery methods and the tuning of {\tt{SExtractor}} \citep{Bertin96} parameters for candidate detection and completeness estimation. We also present our completeness estimates as a function of galaxy properties. In Section 4 we present our final candidates. We compare the results for NGC 628 with theoretical models for the satellite populations in Section 5. Finally, in Section 6 we summarize our results and outline the analysis planned for the full survey. Additionally, tables of {\tt{SExtractor}} parameters and completeness results can be found in the Appendix. 

\section{Description of the LBT-SONG Survey}\label{sec:data}

We use a subset of the data from the ``Survey for Failed Supernovae" \citep{,Kochanek08,Gerke2015,Adams2017}, which consists of UBVR observations of 27 star-forming galaxies taken with the Large Binocular Camera \citep[LBC;][]{Giallongo2008} on the Large Binocular Telescope (LBT). The primary goal of the survey is to discover the formation of black holes by failed core-collapse supernovae, limiting the targets to nearby star forming galaxies. Each galaxy is observed in a single LBC pointing, usually with the target galaxy laying on the central CCD of the camera. Each 2048 x 4096 LBC chip covers 17\parcm3$\times$7\parcm7 with a pixel scale of 0\parcs225, resulting in a total area of 23$^\prime \times \text{23}^\prime$. 

The survey monitors each host in the U, B, and V bands using the LBC/Blue Camera and the R-band on the LBC/Red Camera. The exposure times are scaled with distance to give roughly 1 count per $\text{L}_\odot$ or better for an R band epoch obtained in good conditions. We reduce raw images with overscan correction, bias subtraction, and flat fielding with the \textsc{iraf mscred} package, as described in \citet{Gerke2015}. Image co-addition, astrometric and photometric calibration for the satellite search are described in \citet{garling2019ddo113}, but a shortened version is given here.\par 

Single exposures are astrometrically calibrated using a two-step procedure. First, \texttt{astrometry.net} \citep{Lang2010} is used to obtain an initial astrometric solution, which we then improve using \texttt{SCAMP} \citep{Bertin2006}. We use GAIA-DR2 \citep{GaiaCollaboration2018} as our final astrometric reference, with standard deviations from the reference star positions of $\sim0$\parcs1. These single exposures are then co-added using \texttt{SWarp} \citep{Bertin2002}. The final co-adds have a mean FWHM of $\sim1\parcs0$.

Photometric calibration is done using the \textsc{daophot}, \textsc{allstar}, and \textsc{allframe} PSF fitting photometry packages \citep{Stetson1987,Stetson1994}. After creating our final photometric catalog, we calibrate our photometry by bootstrapping onto SDSS-DR13 \citep{Alam2015}. We used relations from \cite{Jordi2006} to convert SDSS magnitudes to U, B, V, and R with full error propagation. We fit the zeropoints and color terms for all bands simultaneously using an expanded version of the maximum likelihood method described in \cite{Boettcher2013} that accounts for covariances between the zero points and color terms of the four bands. Average calibration uncertainties, including zero point and color term contributions, were 0.03--0.05 mag. Reported magnitudes are corrected for Galactic extinction by interpolating the dust maps from \cite{Schlegel1998} using the updated scaling from \cite{Schlafly11}. 

For NGC 628, the final 5-$\sigma$ point-source depths of the stacked images are 25.0 mag in U and 27.0 mag in the BVR bands. Based on artificial star tests, the 50$\%$ completeness limits are 25.5 mag for U and 27.2 mag for BVR, and the 90$\%$ completeness limits are 24.8 mag in U and 25.8 mag in BVR. 

The fraction of satellite galaxies we expect to observe for a given host depends on three factors: the distance to the galaxy, the stellar mass of the galaxy and the radial distribution of its satellites. Because the hosts range in size from the Small Magellanic Cloud to near Milky Way-mass and span distances from 3 - 10 Mpc, the LBC footprint captures areas ranging from 20 kpc $\times$ 20 kpc to 70 kpc $\times$ 70 kpc. For the hosts in our survey, this corresponds to $\sim0.1\%$ to $\sim20\%$ of the virial volume, based on the $\text{M}_\star-\text{M}_\mathrm{halo}$ relation of \citet{Moster13}. Depending on how satellites are distributed radially, we expect anywhere from 1\%-35\% (if satellites are distributed isothermally) to 5\%-60\% (if satellites trace the smooth halo of the host) of the bound satellites to lie within the LBC footprint. The two choices of radial distribution bracket the extremes typically considered in the literature, and reflect the theoretical and observational uncertainties \citep{Dooley2017a,Kim2018}.

NGC 628, shown in Figure \ref{fig:Host}, is an isolated (tidal index $\Omega_{1} = -0.3$ \citealt{Karachentsev2013updated}) face-on spiral host with an estimated stellar mass of $\sim1.3\times 10^{10} \text{M}_\odot$ \citep{Leroy2008THINGS}, roughly 1/4 that of the MW. Assuming the $\text{M}_\star-\text{M}_{\mathrm{halo}}$ relation of \citet{Moster13}, we estimate a halo mass of $\text{M}_{\mathrm{halo}} \approx (3-4)\times 10^{11} \text{M}_\odot$, with a virial radius of approximately 200 kpc, based on the overdensity criterion of \citet{Bryan98}. 
We use the TRGB distance estimate of 9.77 Mpc from \citet{Mcquinn2017Distance}, for which a 200 kpc virial radius corresponds to 70$^\prime$. The LBT footprint for NGC 628 corresponds to 65 kpc $\times$ 65 kpc. We expect $\sim 5\%-40\%$ of NGC 628's satellites to lie within the LBC footprint, depending on the radial distribution of satellites. If the satellites have the same radial distribution in NGC 628 as the classical satellites of the Milky Way, we expect the fraction to be $\sim 25\%$ \citep{Dooley17b}.

\section{The Diffuse Dwarf Galaxy Detection Pipeline} \label{Pipeline}
In this section, we describe how we search for dwarf galaxies as diffuse, extended sources in the case of distant dwarfs (distance > 5 Mpc) in which individual component stars are not resolved. Many previous works to detect galaxies in the diffuse regime have field has used visual searches \citep[e.g.,][]{Kim2011, Park2017, muller2017m, smercina2018lonely, Muller2018a, crnojevic2019faint}. More recently, semi-automated pipelines employing  {\tt{SExtractor}} have been done to search for diffuse and low surface brightness galaxies over large survey areas \citep{van_der_Burg_2016,Bennet_2017,Greco18,Carlsten_2020}. Many of these studies inject mock galaxies at random locations in the data to quantify survey completeness. These searches are typically over wide fields and search for objects with large angular size, employing size cuts on candidates that are usually large to cut down on contamination, therefore prioritizing purity over completeness. The LBT-SONG survey region is smaller and centered on the hosts, leading to two design principles that distinguish the LBT-SONG diffuse galaxy search pipeline from other mock galaxy injection searches: 1) Our background varies on account of the stellar halo of the host, in addition to the usual cirrus, and we attempt to quantify that, and 2) We are not as restrictive on size cuts, using the host distance to help design size cuts, which allows us to find even relatively small dwarf galaxies, as expected by the size-luminosity relation of Local Group dwarf galaxies. Therefore, the LBT-SONG diffuse galaxy search pipeline described here creates and injects mock galaxies with a range of properties systematically, throughout the entire survey region to quantify completeness as a function of galaxy model, and therefore size, and position.

\subsection{Initial visual inspection}
The initial visual inspection of the data revealed a large, bright satellite candidate, NGC 628 dwA (see Figure \ref{fig:Host}), which appears visually `lumpy' within its central region, rather than smooth, which we would expect to see if
it is at the distance of the host (10 Mpc), and not a background galaxy. This candidate is in the $\text{M}_{\text{V}}$ range ($\text{M}_{\text{V}}<-10$) to which we are complete via visual inspection. To robustly detect and quantify completeness in the fainter diffuse dwarf regime, we employ the diffuse dwarf galaxy detection pipeline.

\subsection{The pipeline}
In what follows, we present a quantitative detection and completeness estimation method for dwarf satellite candidates with $-6>\text{M}_{\text{V}}>-9$. 
We use {\tt{SExtractor}} to recover mock galaxies and identify potential candidates in the images. {\tt{SExtractor}} allows for a broad range of parameter choices. To determine the optimal detection parameters, we begin by injecting mock galaxies into the images and optimizing the {\tt{SExtractor}} parameters to recover them. In this way we can optimize our detection while at the same time automatically understanding our completeness. The optimal parameters are then used for our candidate search. We first outline our procedure and then describe it in detail.

\begin{enumerate}
    \item \textbf{Simulate mock galaxies} \\
    We simulate 54 simple-stellar-population (SSP) galaxy models using Plummer profiles \citep{Plummer1911} with luminosities, surface brightnesses, ages and metallicities spanning the range of properties found among Local Group dwarfs, as summarized in Table \ref{table:mocks}. 
	\item \textbf{Inject the mock galaxies into the images} \\
	The mock galaxy models are injected into the images on grids spanning the field, avoiding overcrowding of the mock galaxies or injecting too much flux and altering the background. The grid ensures that the mock galaxies fully cover the image footprint, as required to calculate the completeness. 
	\item \textbf{Create weight images and masks} \\
	We create a mask of bright foreground and background objects using the iterative-thresholding methods of \citet{Greco18}. We inflate the masked regions and superimpose them onto the weight maps which are used to account for the large background variations across the images due to the presence of NGC 628.
	\item \textbf{Determine {\tt{SExtractor}} search parameters} \\
	For each mock galaxy, we run {\tt{SExtractor}} multiple times, varying the DETECT{\_}MINAREA and ANALYSIS{\_}THRESH parameters and the detection filter. We select the detection filter and parameters which optimize the ratio between the number of detected mock galaxies and the total number of detections. In other words, we choose those parameters and detection filters which maximize the completeness while minimizing the number of false positives. The parameters are optimised for each type of mock galaxy individually. These parameters are used for the final runs to search for candidate galaxies and to characterize the completeness.
	\item \textbf{Apply selection cuts based on {\tt{SExtractor}} output of recovered mock galaxies} \label{scuts}\\
    We run {\tt{SExtractor}} with the optimized parameters obtained in the previous step on all grids of mock galaxies and develop selection cuts to better discriminate the real mock galaxies from false positives. These additional cuts are included before calculating the completeness. The cuts are summarized in Table \ref{table:cuts}. 
    \item \textbf{Apply the same procedures to the real data} \\
    We run 54 instances of {\tt{SExtractor}} on the real data to obtain the initial list of candidates. Each {\tt{SExtractor}} run uses the optimized parameters and detection filters determined for each galaxy model in the previous steps. We apply additional selection cuts from step \ref{scuts}, to obtain a list of candidates for visual inspection.
	\item \textbf{Visual inspection} \\
	We visually inspect the $\sim 600$ remaining candidates, removing any obvious false positives. These false positives are dominated by blends of point-like sources with background and foreground diffuse light (e.g., a nearby bright star, spiral arms of the host, massive low-z galaxies, or Galactic cirrus). Several examples are shown in Fig.~\ref{fig:rejects}. Once this is complete, we are left with the candidate NGC 628 dwB. 
\end{enumerate}

\subsection{Mock galaxies} \label{subsec:mocks}

We simulate a total of 54 artificial galaxies using SSP models for the stars. Our framework is similar to that of ArtPop \citep{danieli2018}. We attempt to capture the range of properties of Local Group dwarf galaxies and therefore create mock galaxies varying the following:

\begin{enumerate}
    \item \textbf{Absolute V band magnitude} \\
    The depth of our data permits us to probe integrated dwarf galaxy magnitudes down to $\text{M}_{\text{V}} \sim -6$. We increase the absolute V band magnitude in increments of 1 mag up to $\text{M}_{\text{V}}  = -9$, beyond which we are visually complete to satellites in unmasked regions. 
    \item \textbf{Average surface brightness within the half light radius} \\  
    We vary the surface brightness from 25 \surfb{}, corresponding to the highest surface brightness Local Group dwarfs, down to 28 \surfb{}, our empirically-determined surface brightness limit, in increments of 1 \surfb{}. Nearly two-thirds of Local Group galaxies with magnitudes between $\text{M}_{\text{V}}  = -9$ and $-6$ have effective surface brightnesses between $\mu_V$ = $25 - 28 \hbox{ mag}/\Box^{\prime\prime}$ \citep{McConnachie12}.
    \item \textbf{Age} \\
    We simulate 3 ages: 100 Myr, 1 Gyr and 10 Gyr to span the range of star formation histories found among Local Group dwarfs \citep{weisz2014b}.
    \item \textbf{Metallicity} \\
    We choose metallicities of 0.1 $\text{Z}_{\odot}$ and 0.01 $\text{Z}_{\odot}$ which are typical of Local Group dwarfs \citep{Kirby13}.  
\end{enumerate}

\begin{table}
\centering
 \caption{The range of simulated dwarf galaxy properties.}
 \label{table:mocks}
 \begin{tabular}{ccc}
  \hline
  Property & Values \\
  \hline
  $\mu_{\rm eff}$ in V band ($\rm mag/\square''$)& 25, 26, 27, 28 \\
  $\text{M}_{\text{V}}$ & $-6$,$-7$,$-8$,$-9$ \\
  $\left[ \text{Fe} \textfractionsolidus \text{H} \right]$ & $-2$,$-1$ \\
  Age (Myr) & $10^2, 10^3, 10^4$ \\
  \hline
 \end{tabular}

\end{table}

\begin{figure}
 \includegraphics[width=\columnwidth]{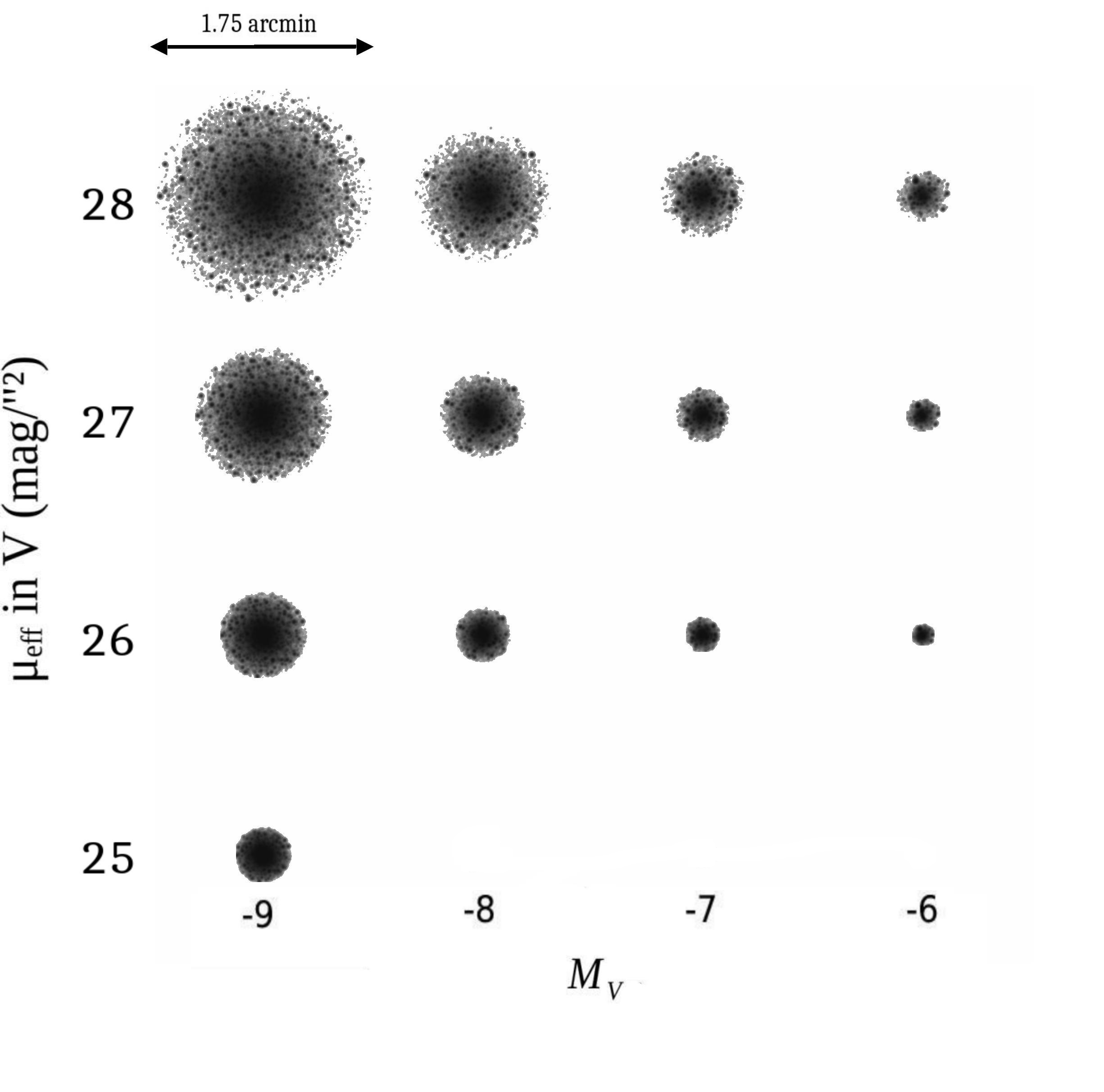}
 \caption{A gallery of sixteen 10 Gyr, 0.01 $\text{Z}_\odot$ galaxy models. The galaxies with $\text{M}_{\text{V}}=-8$,$-7$,$-6$ and $\mu_{\text{V}}=$25 \surfb{} are not modeled as they are visually indistinguishable from background sources.}
 \label{fig:gallery}
\end{figure}

The SSP models are derived using the \citet{Marigo17} isochrones through the CMD v3.1 web tool\footnote{\url{http://stev.oapd.inaf.it/cgi-bin/cmd}} using the LBT/LBC filter set. For each SSP galaxy model, we sample a stellar distribution with a \citep{Chabrier2001} initial mass function (IMF) to generate a set of stars down to $\text{M}_{\text{V}} = +6$. We model the light distribution as a Plummer sphere based on models of Milky Way dwarfs \citep{Munoz18}, with the half-light radius determined by the model's total luminosity and average surface brightness. Stars are reddened according to the \citet{Schlafly11} extinction map. Finally, the simulated galaxies are convolved with an empirical PSF derived from stars selected from the final image stacks using {\tt{SExtractor}}, and Poisson noise is added. We considered mock galaxies with ellipticities that were: zero ($e=0$), moderate ($e=0.35$ based on the Milky Way satellites' average ellipticity; \citet{McConnachie12}), or high ($e=0.83$, the highest ellipticity of any Milky Way satellite; \citet{McConnachie12}). We report the completeness results for the zero ellipticity case, as the completeness measurements for the higher ellipticity cases were within 5\% of those from the zero eccentricity case. 

\subsection{Injecting mock galaxies}
\begin{figure}[H]
 \includegraphics[width=3.2in]{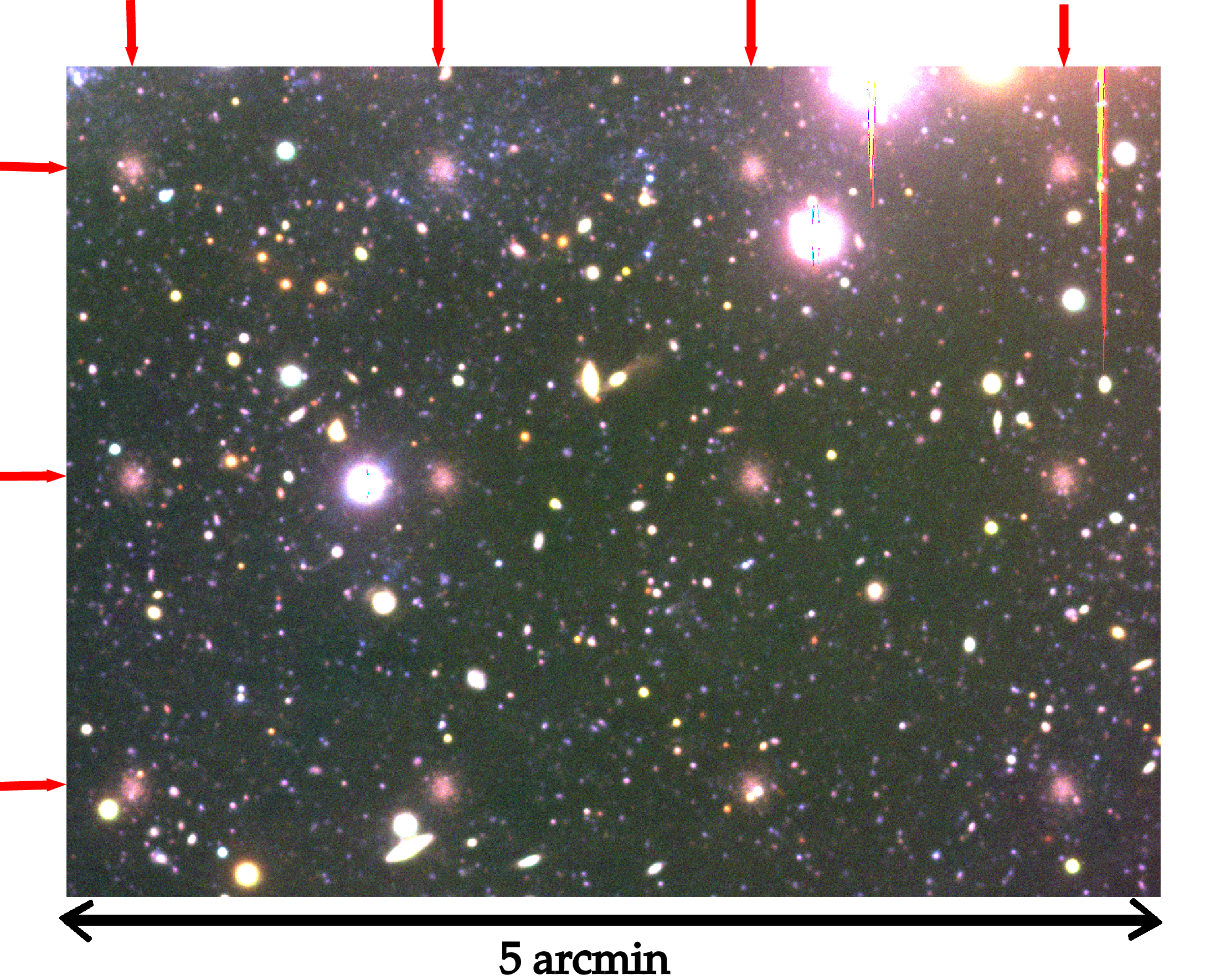}
 \caption{A section of a composite color RVB stacked image with a grid of the 10 Gyr, 0.01 $\text{Z}_{\odot}$, $\text{M}_{\text{V}}=-9$, $\mu_{\text{V}}=$ 26 \surfb{} dwarf galaxy model injected at an assumed distance of 9.77 Mpc. The red arrows indicate the lines of the grid.}
 \label{fig:grid}
\end{figure}

We wish to measure our ability to detect each type of dwarf generated in the previous subsection separately. To achieve this, we place a grid of simulated dwarfs of each type in the real data assuming a host distance of 9.77 Mpc. An example is shown in Figure \ref{fig:grid}. We create multiple images with grid positions that are offset from one another. Using multiple, offset grids of the same mock galaxy, we are able to effectively place a mock galaxy throughout the entire image, which we require for robust completeness measurements. Simultaneously, the mock satellites are well-separated in each individual image and sufficiently small in number to leave the estimated backgrounds unchanged. The number of mock galaxies injected are therefore based on their size and range from 920 per chip ($M_{V}$ = $-9$, 28 \surfb{}) to 14740 per chip ($M_{V}$ = $-7$, 26 \surfb{}).

\subsection{Masking} \label{masking}
We create two masks: one for the 10 Gyr mock galaxies, with a total of $29\%$ of the area masked, and a more conservative $60\%$ mask for the 100 Myr and 1 Gyr models in which large regions dominated by the host's star-forming regions are additionally masked. This is necessary because these younger stellar populations are more challenging to distinguish from star forming regions in the host.

We first create {\tt{SExtractor}} weight images to account for the spatially varying background from the host's spiral arms, Galactic cirrus, and foreground contamination from Milky Way stars. We masked high surface brightness sources in R band and their associated diffuse light using the iterative thresholding technique from \citet{Greco18}. These masks are then extended by 10 pixels to mask more of the diffuse light and applied to the weight images used for {\tt{SExtractor}}, resulting in the final masks for the detection of old (10 Gyr) models.

The second mask, which is used when searching for dwarfs with younger stellar populations, is created by using the mask for older mock galaxies as a starting point and additionally masking large star-forming regions of the host by hand. These significantly larger masks drive the lower completeness for young model galaxies. 

\subsection{{\tt{SExtractor}} parameters search} \label{SE param search}
Having generated a grid of mock candidate dwarfs, we now wish to determine the best possible set of parameters for detecting those dwarfs. For each of the different types of dwarfs, we ran {\tt{SExtractor}} over a grid of values for DETECT{\_}MINAREA and ANALYSIS{\_}THRESH parameters and each of the filters as detection band.

Next, we compare each output catalog of detected objects with the known input objects. We find the center of the closest {\tt{SExtractor}} detection within 10-30 pixels, depending on the size of the injected galaxy, to the position of the injected galaxies and count that as a recovered galaxy. We avoid combinations of parameters that result in random noise being detected throughout the image and being misidentified as a `recovered galaxy' by setting the minimum possible DETECT{\_}THRESH = 0.5. The optimal DETECT{\_}MINAREA and DETECT{\_}THRESH grid values depend on the size and surface brightness of the target galaxy. We choose the tophat{\_}1.5{\_}3x3.conv filter and set DEBLEND{\_}MINCONT = 0.1, ANALYSIS{\_}THRESH = 1 and DEBLEND{\_}NTHRESH = 64. These values were chosen to minimize the fragmentation of the larger diffuse galaxies into multiple {\tt{SExtractor}} detections. For each dwarf model, we selected the parameters which simultaneously found at least $90\%$ of the injected models, and had the smallest total number of {\tt{SExtractor}} detections. These final parameters are used both for the final search and to compute our completeness. The final {\tt{SExtractor}} parameters are given in Table 3 in the Appendix.

\subsection{Selection cuts on {\tt{SExtractor}} output of mock galaxies }

\begin{table}
\centering
 \begin{tabular}{ccc}
  \hline
  Selection cut & mock galaxies & candidates\\
  \hline
  original & 634210 & 6091\\
  $<25\%$ of $\text{r}_{0.5}$ masked & 593912 & 5178 \\
  R<20.5 mag& 591856 & 5151\\
  $\text{r}_{0.5}$ > 6 pixels & 552714 & 1310 \\
  $\text{r}_{0.5}$ < 45 pixels & 552501  & 1299 \\
  $\text{r}_{0.1}$ > 2.2 pixels & 550447 & 1192 \\
  $\text{r}_{0.1}/\text{r}_{0.5}$>0.21 & 549566 & 1161 \\
  $\text{r}_{0.1}/\text{r}_{0.5}$<0.6 & 549276 & 989 \\
  SE flag < 3 & 468237 & 573 \\
  \hline
 \end{tabular}
 \caption{A summary of all selection cuts and remaining recovered mock galaxies and candidates after each consecutive cut.}
 \label{table:cuts}
\end{table}

We next compare the distributions of various {\tt{SExtractor}} output parameters for the mock galaxies and false positive detections to develop a series of selection cuts. A summary of our final selection cuts is given in Table \ref{table:cuts}. 

We first removed mock galaxies with R < 20.5. These objects were generally bright foreground stars or distant background galaxies. If $\text{r}_{0.5}$ and $\text{r}_{0.1}$ are the {\tt{SExtractor}} radii containing 50\% and 10\% of the light, respectively, we remove objects with $\text{r}_{0.5} < 6$ pixels (1\parcs35), $\text{r}_{0.1} < 2.2$ pixels, $\text{r}_{0.5} > 45$ pixels and objects outside the range $0.21 < \text{r}_{0.1}/\text{r}_{0.5} < 0.6$. Figure \ref{fig:hist} shows an example for why we set the requirement of $\text{r}_{0.5}>$ 6 pixels. Objects with $\text{r}_{0.5}$, $\text{r}_{0.1}$ or $\text{r}_{0.1}/ \text{r}_{0.5}$ too small are generally host halo stars or background galaxies. Objects with $\text{r}_{0.5}$, $\text{r}_{0.1}$ or $\text{r}_{0.1}/ \text{r}_{0.5}$ too large are generally cirrus or lower surface brightness emission near masked regions misidentified as $\mu_{\text{V}}=$ 28 \surfb{} mock galaxies. Lastly, we removed objects with {\tt{SExtractor}} flags of 3 or larger which are objects that are both deblended and have more than $10\%$ of bad pixels in the aperture, or have saturated pixels or are too close to the image edge. We examined adding color-selection, but background variations and other contamination problems made it unworkable. 

\begin{figure}[H]
 \includegraphics[width=3.2in]{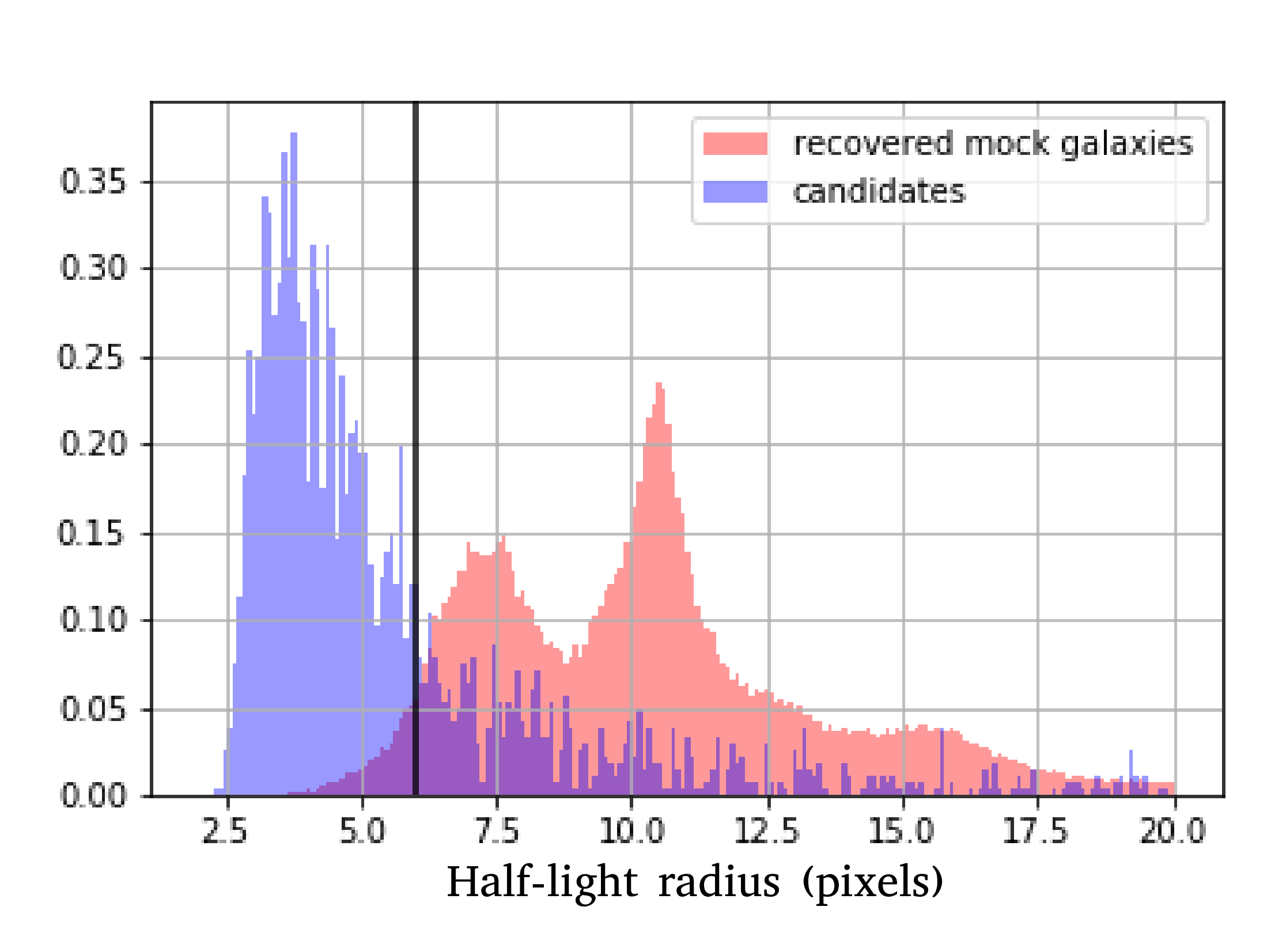}
 \caption{A normalized histogram of the {\tt{SExtractor}} half-light radii of all recovered mock galaxies for NGC 628 in red and all the dwarf satellite candidates in blue. The black line shows the selection cut made at 6 pixels. This was one of the most effective cuts.}
 \label{fig:hist}
\end{figure}

\subsection{Applying the procedure to the real data}

We run all sets of {\tt{SExtractor}} parameters and detection filters optimized for the mock galaxies on the real data and apply the same selection cuts used on the mock galaxies to obtain the candidate list for visual inspection. Before applying the selection cuts, we started with 6091 candidates. We expect only a small number of dwarf satellites in the footprint, therefore the vast majority of these objects are spurious detections. With the cuts, we remove only $26\%$ of the mock galaxies, but $91\%$ of the candidates. The selection cuts based on size were the most effective at removing candidates but not mock galaxies. Figure 4 shows the distribution of estimated half-light radii, $\text{r}_{0.5}$, for the injected dwarfs and all detected objects.

\begin{figure*}[!b]
\begin{centering}
 \includegraphics[width=5.0in]{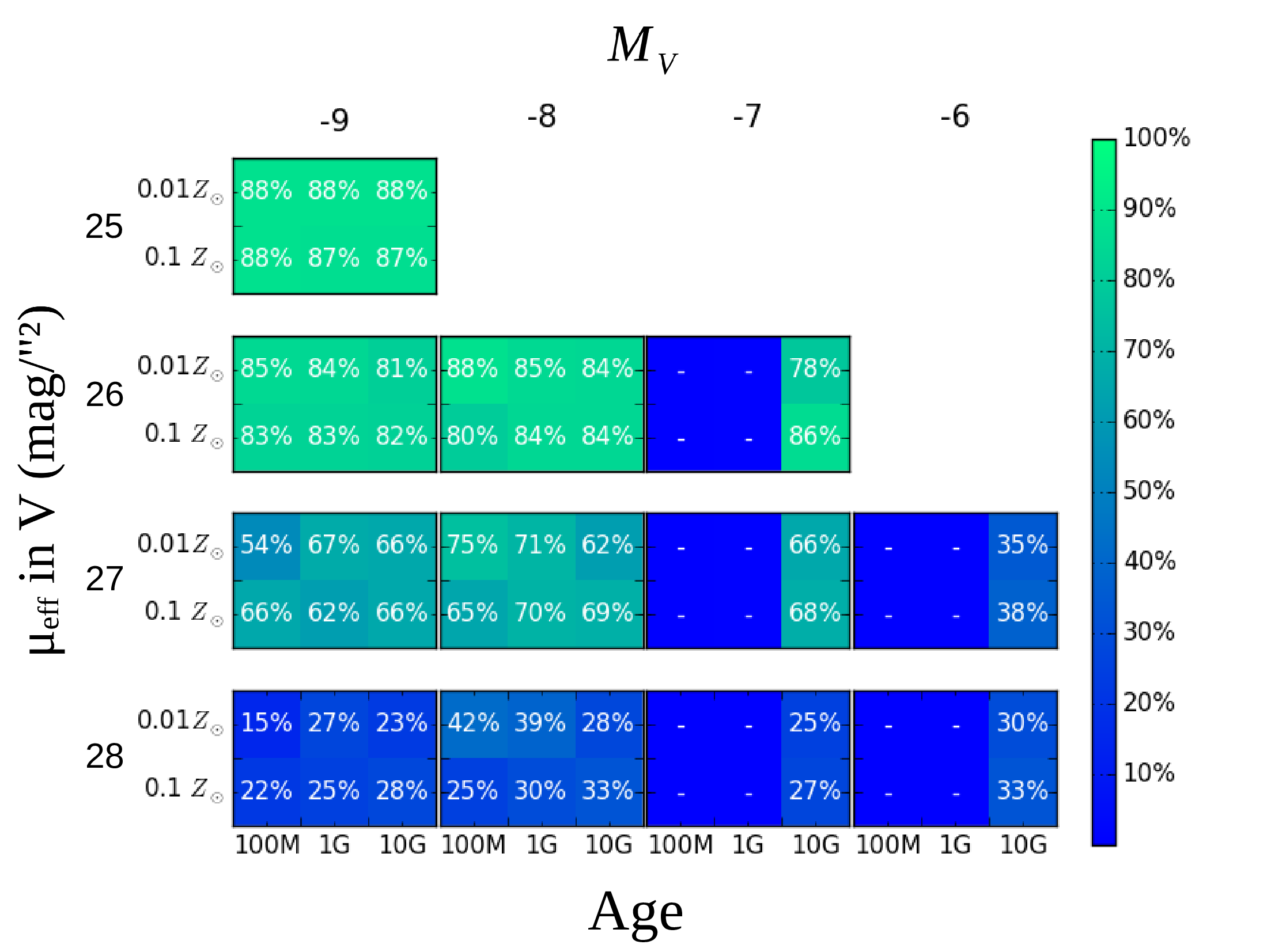}
 \setcounter{figure}{5}
 \caption{Completeness in the unmasked regions for all mock galaxy models. Each boxed region represent models of a single $\rm \text{M}_{\text{V}}$ and $\mu_{\text{V}}$, increasing in V band magnitude towards the right and decreasing in surface brightness downward. Within each box are 6 numbers representing the completeness result for a model of that $\rm \text{M}_{\text{V}}$ and surface brightness with the age and metallicity given by its column and row respectively, within each box, as labeled along the bottom and left of the figure. The boxes with a '-' were not simulated.}
 \label{fig:complAdjusted}
\end{centering}
\end{figure*}

After applying the cuts, 573 candidates remain. After visual inspection, we identify only one convincing candidate, NGC 628 dwB. Examples of rejected candidates are shown in Figure \ref{fig:rejects}. NGC 628 dwB was detected by 29 distinct sets of the {\tt{SExtractor}} parameters out of a total 54. This included parameters tuned for the faint and old models (age = 10 Gyr, M$_{\text{V}}= -6, -7$) and those tuned for the most luminous and young models (age = 100 Myr, M$_{\text{V}}= -8, -9$). NGC 628 dwB, together with NGC 628 dwA, which was discovered in the initial visual inspection, constitute the final candidates. 

\begin{figure}[H]
 \includegraphics[width=3.2in]{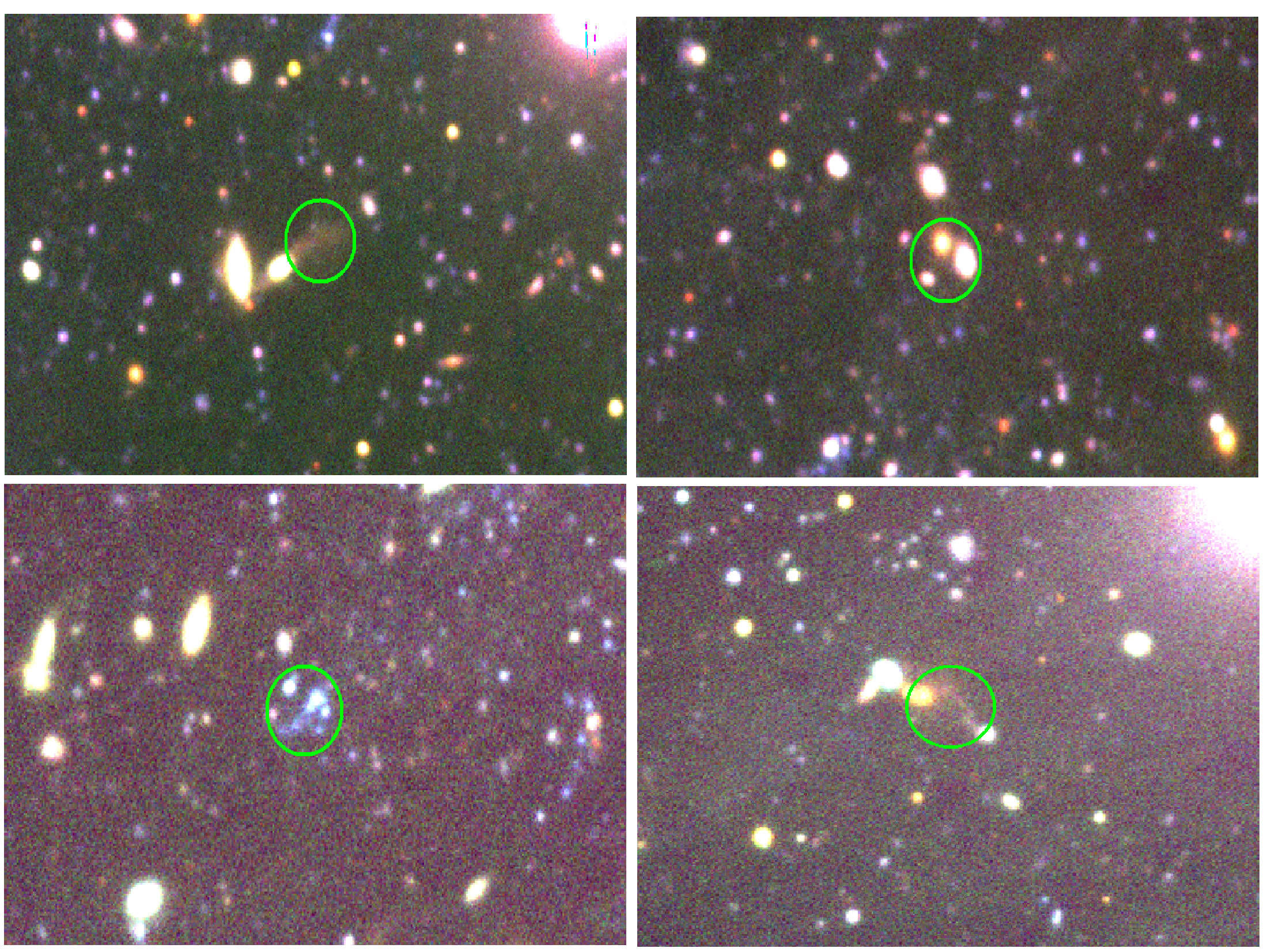}
 \setcounter{figure}{4}
 \caption{A sample of candidates that were rejected by visual inspection. Top left: a tidal tail. Top right: an over-density of background galaxies or stars. Bottom left: a cluster of young stars near the host. Bottom right: diffuse light from interacting galaxies.}
 \label{fig:rejects}
\end{figure}

\begin{figure*}[!t]
\begin{centering}
\setcounter{figure}{6}
 \includegraphics[width=5.0in]{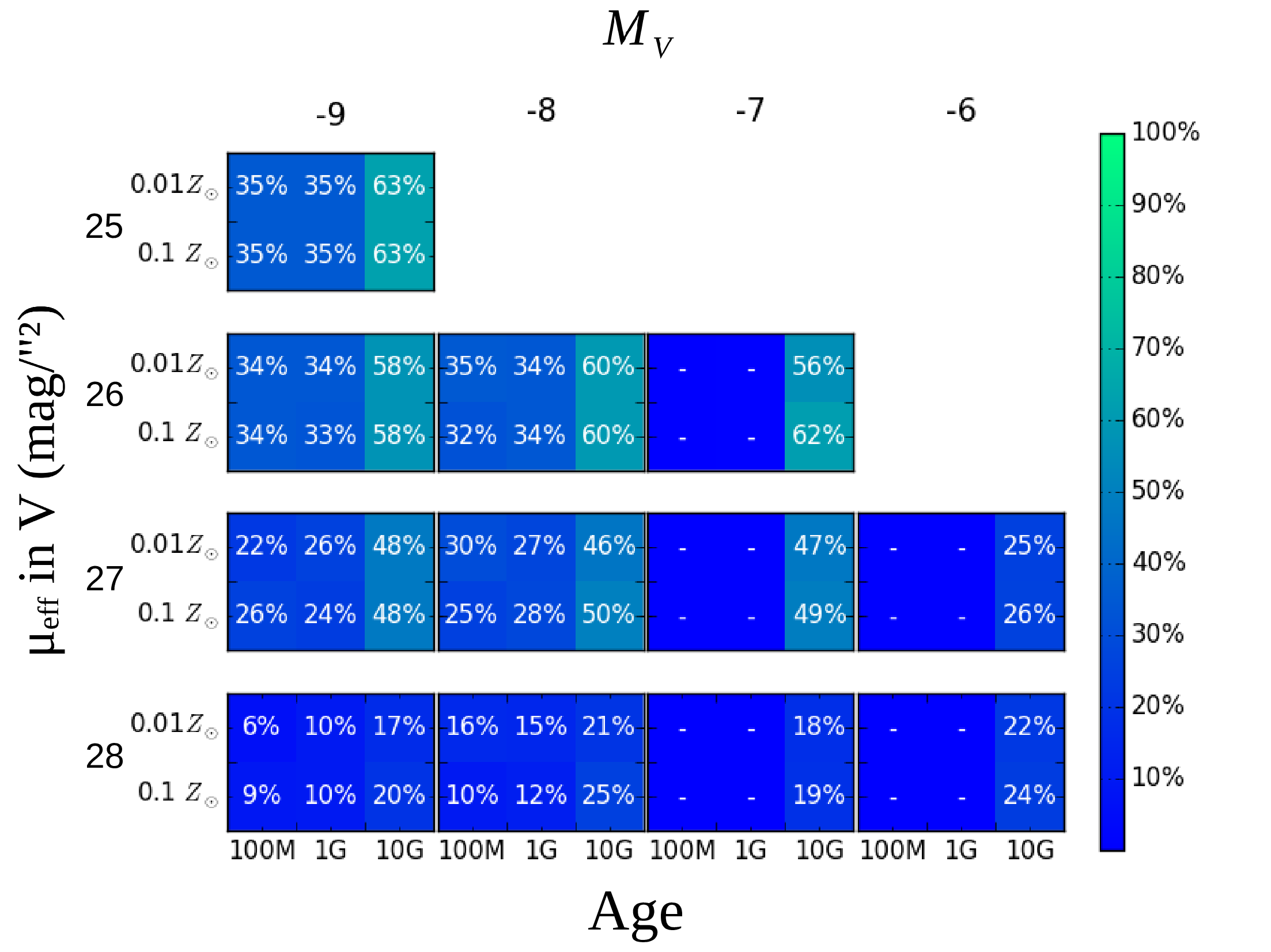}
 \caption{Completeness after correcting for the area lost due to masking. The structure of the figure is the same as for Figure \ref{fig:complAdjusted}.}
 \label{fig:complRaw}
\end{centering}
\end{figure*}

\subsection{Completeness Results}\label{sec:completeness}

In order to derive our luminosity function for NGC 628, we need to know our completeness, which we compute for every individual model and illustrate in Figures \ref{fig:complAdjusted} and \ref{fig:complRaw}. These percentages are the number of recovered galaxies that passed all selection cuts, divided by the total number of injected galaxies. Figure \ref{fig:complAdjusted} shows the completeness for the unmasked search area, while Figure \ref{fig:complRaw} shows the reduced completeness after correcting for the search area lost due to masking. In both figures, there is a box for each model in $\rm \text{M}_{\text{V}}$ and $\mu_{\text{eff}}$, increasing in V band magnitude towards the right and decreasing in surface brightness downward. In each box there are 6 numbers for the completeness as a function of age and metallicity.

As expected, we are more complete for the brighter, higher surface brightness galaxy models, namely those with $\text{M}_{\text{V}} = -9$ and $\mu_{\text{eff}}$ = 25, 26, or 27 \surfb{}, and $\text{M}_{\text{V}} = -8$ with $\mu_{\text{eff}}$ = 26, or 27 \surfb{}. We become less complete with increasing $\text{M}_{\text{V}}$ and increasing $\mu_{\text{eff}}$. Models with $\text{M}_{\text{V}} = -6,-7$ are easily detected but will more often have measured half-light radii below our selection cut of 6 pixels. Models with more diffuse central regions, $\mu_{\text{eff}}$ = 27, 28 \surfb{}, become increasingly difficult for {\tt{SExtractor}} to detect at this distance.

The completeness calculations show that we recover about two-thirds of the artificial galaxies for a fixed model, largely missing satellites due to masking (Fig.~\ref{fig:complRaw} vs. Fig.~\ref{fig:complAdjusted}). In unmasked regions, we are relatively complete for older stellar populations that are not too low or too high in surface brightness. Many, but not all, Local Group satellites down to $\text{M}_\text{V} = -6$ have structural and stellar population properties similar to these models  (Fig.~\ref{fig:MWDwarfs}). However, at fixed stellar mass, the distribution of galaxy size, luminosity, and star-formation histories are highly uncertain \citep[see ][ for a discussion]{danieli2018}. Our best estimate is that about 50\% of satellites in the LBC footprint and in the magnitude and surface brightness ranges shown in Fig.~\ref{fig:complAdjusted} are detectable by our methods. 

\section{Dwarf galaxy candidates}
\begin{figure*}[!b]
 \includegraphics[width=\textwidth]{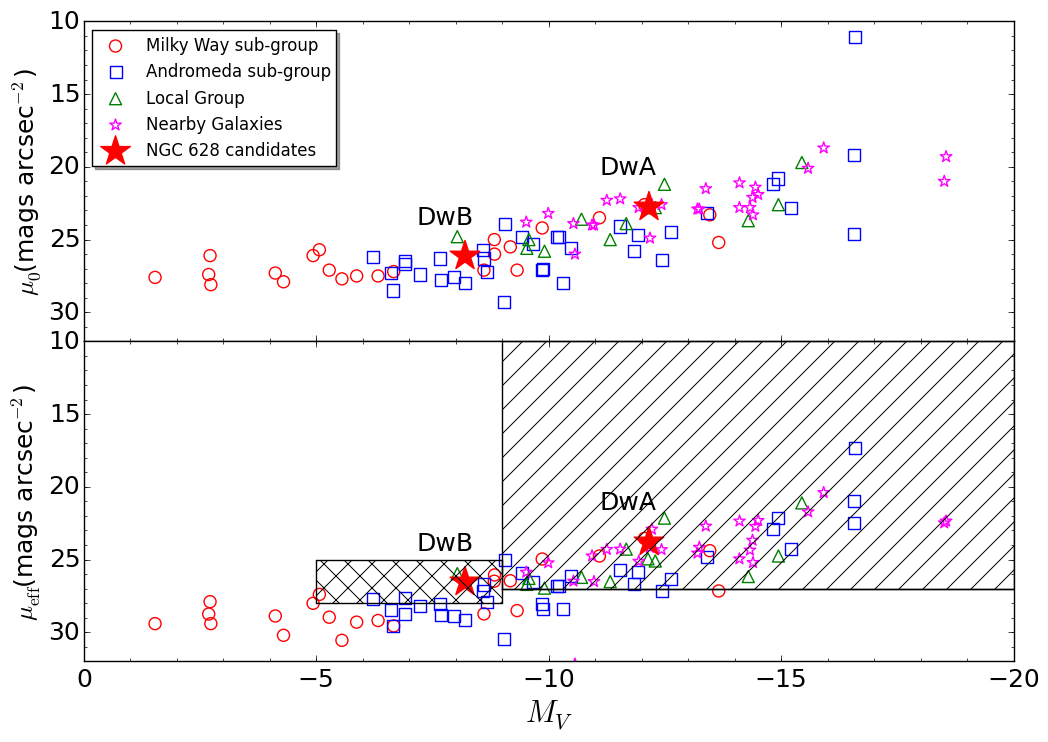}
   \setcounter{figure}{8}
 \caption{The V band surface brightness versus $\text{M}_{\text{V}}$ for NGC 628 dwA and dwB at an assumed distance of the host are shown as red stars. We compare them to the galaxy sample from \citet{McConnachie12} using updated values. The symbol colors identify satellites of the Milky Way, M31, Local Group and nearby galaxies. The top panel uses the central surface brightness, and the bottom panel uses the average surface brightness interior to the half-light radius. The smaller hashed region in the bottom panel represents the parameter space of our mock galaxies which the pipeline is sensitive to, while the larger hashed region represents the region we are visually complete to.}
 \label{fig:MWDwarfs}
\end{figure*}

The two satellite galaxy candidates, NGC 628 dwA and NGC 628 dwB, are shown in relation to the host in Figure \ref{fig:Host}. Color cutouts are presented in the left panels of Figure \ref{fig:GalfitPics}. Their positions, along with the photometric and structural properties we measure, and data from other surveys which we discuss below, are summarized in Table \ref{table:candresults}. Both candidates appear visually `lumpy' within their central regions, rather than smooth, which we would expect to see if they were at the distance of the host (10 Mpc). If they are satellites, they are both $\sim$35 kpc in projection from NGC 628. 

\begin{figure*}[!b]
 \includegraphics[width=\textwidth]{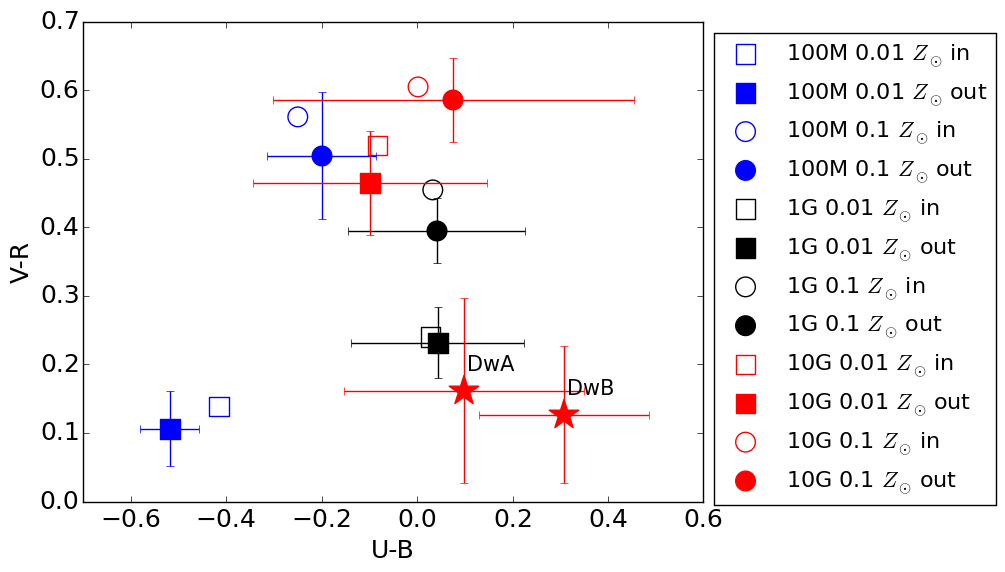}
 \caption{The V$-$R and U$-$B colors of the galaxy models with $\text{M}_{\text{V}}=-8$ and $\mu_{\text{V}} = 27 \hbox{ mag}/\square''$ for ages of 100 Myr, 1 Gyr, 10 Gyr and metallicities of 0.1 $\text{Z}_\odot$ and 0.01 $\text{Z}_\odot$. The filled circles are averages of the measured colors for 20 galaxies of each model, injected near dwB. Error bars are the standard deviation in the color measurement. {\tt{SExtractor}} was run with a 10 pixel fixed aperture for all objects to obtain magnitudes. The open circles are the {\tt{SExtractor}} colors obtained from running the software on those same models prior to injection into the stacked images. The two red stars are our two candidates, with uncertainties propagated from color uncertainties in Table \ref{table:candresults}.}
 \label{fig:color-color}
\end{figure*}

We use GALFITM \citep{GALFITM} which models all bands simultaneously and is based on GALFIT \citep{GALFIT}, to estimate the structural and photometric properties of our candidates (Table \ref{table:candresults}). The magnitudes are Galactic extinction corrected. The best-fit models are shown in Figure \ref{fig:GalfitPics}. We estimate uncertainties on the \textsc{galfitm} results by producing 100 analogs of dwA and dwB star by star, as described in \S 3.1 of \citet{garling2019ddo113}. The stellar positions are sampled from 2D S\`ersic profiles with the morphological parameters measured by \textsc{galfitm}, while the stellar magnitudes are sampled from PARSEC 1.2S isochrones \citep{Aringer2009,Bressan2012,Chen2014} including the thermally-pulsating asymptotic giant branch and other improvements from \citet{Marigo17}. We use the \cite{Chabrier2001} lognormal IMF and add the Galactic extinction derived in \S 2 to the stellar magnitudes. Stars are added to the mock galaxies iteratively until their apparent magnitudes in V band match those measured by \textsc{galfitm}. When all star positions and magnitudes have been generated, they are injected into the LBT images by \textsc{addstar}, using the \textsc{daophot} PSFs. We then run \textsc{galfitm} on these analogs the same way as on the real galaxies. We derive approximate 1-$\sigma$ uncertainties on fitted quantities by comparing the \textsc{galfitm} results to the true values for 100 analogs of each of dwA and dwB. We find uncertainties on the magnitudes are typically 50\% higher than those reported by \textsc{galfitm}, likely due to the galaxies being semi-resolved.

\begin{figure}[H]
 \includegraphics[width=3.2in]{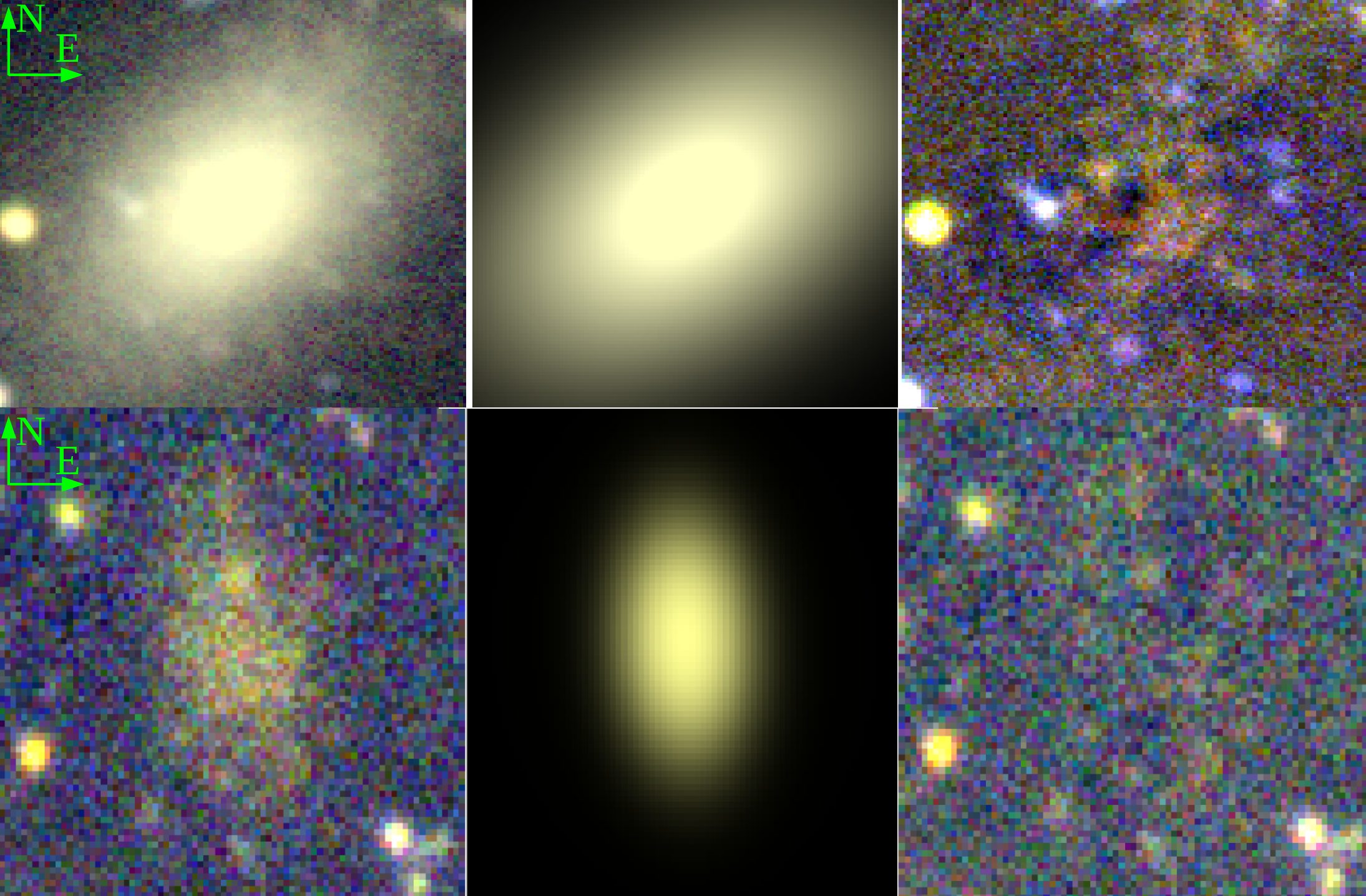}
  \setcounter{figure}{7}
 \caption{The two dwarf galaxy candidates: NGC 628 dwA (top) and NGC 628 dwB (bottom). From left to right: composite BVR color images of the candidates, a composite BVR GALFITM model, and a composite of the subtraction residuals.}
 \label{fig:GalfitPics}
\end{figure}

\begin{table}
\centering
 \caption{Properties of the satellite candidates.}
 \label{table:candresults}
 \begin{tabular}{ ccc p{1cm} }
  \hline
  Candidate & NGC 628 dwA & NGC 628 dwB \\
  \hline
  RA (J2000) & 1:37:17.8 & 1:36:23.2\\
  Dec (J 2000) &  +15:37:58.2 & +15:57:53.0\\
  Separation ($^\prime$) & 12.6 & 11.8\\
  Separation (kpc) & 36 & 34 \\
  $\text{m}_{\text{U}}$ & 17.49 $\pm$ 0.15 & 22.52 $\pm$ 0.25\\
  $\text{m}_{\text{B}}$ & 18.34 $\pm$ 0.09 & 22.67 $\pm$ 0.14\\
  $\text{m}_{\text{V}}$ & 17.80 $\pm$ 0.07 & 22.23 $\pm$ 0.13\\
  $\text{m}_{\text{R}}$ & 17.64 $\pm$ 0.06 & 22.12 $\pm$ 0.10\\
  radius (pixels) & 34.64 $\pm$ 0.81 & 15.57 $\pm$ 0.75 \\
  radius (arcsec) & 7.79 $\pm$ 0.07 & 3.50 $\pm$ 0.02 \\
  S\'ersic index & 0.95 $\pm$ 0.02 & 0.39 $\pm$ 0.08 \\
  axis ratio & 0.64 $\pm$ 0.01 & 0.58 $\pm$ 0.03\\
  W1 & 16.33 $\pm 0.065$& >16.6 \\
  W2 & 16.16 $\pm 0.198$& >15.6 \\
  W3 & >12.63 & >11.3 \\
  W4 & >9.05 & >8.0 \\
  GALEX NUV & 21.66 $\pm 0.27$ & >22.7 \\
  GALEX FUV & >22.18 & >22.7\\
  \hline
 \end{tabular}
 \captionsetup{labelformat=empty}
 \caption{The magnitudes are colorterm-calibrated and Galactic extinction corrected, and their errors are derived from the analog mock galaxy insertion tests and include calibration uncertainties. Errors for the physical parameters are those output by GALFIT. Infrared and NUV data are from the AllWISE Data Release and GALEX NGS. The upper limits for dwA are calculated at the location of the candidate. Upper limits for dwB are the 5-$\sigma$ point source limits from ALLWISE \citep{ALLWISE,WISE2010A} and the GALEX NGS \citep{GALEX2005} surveys.}
\end{table}

We can also make comparisons with Local Group satellites. Assuming both candidates are at $\sim10$ Mpc, dwA has $\text{M}_{\text{V}} = -12.2$ and dwB has $\text{M}_{\text{V}} = -7.7$. This makes dwA a classical dwarf and puts dwB in the ultrafaint regime, in terms of luminosity. The central and effective surface brightness and absolute V band magnitudes for both candidates are shown in Figure \ref{fig:MWDwarfs} along with other known Local Group and Local Volume dwarfs. The structural properties of both candidate dwarfs are consistent with those populations.

To estimate the ages of the candidates, we examine their colors. Figure \ref{fig:color-color} shows the $\text{V}-\text{R}$ and $\text{U}-\text{B}$ colors of the candidates along with those of the mock galaxies with morphological properties most similar to dwB, namely those models with $\text{M}_{\text{V}} = -8$ and average central surface brightness of 27 \surfb{}. We show all the trial ages of (100 Myr, 1 Gyr, 10 Gyr) and metallicities (0.1  $\text{Z}_{\odot}$ and 0.01 $\text{Z}_\odot$). For each galaxy model, we measure the colors of 20 mock galaxies injected near the position of dwB. We find that both candidates are most similar in color to mock galaxies of intermediate age (1 Gyr) and Z = 0.1 $\text{Z}_{\odot}$.

To obtain mass estimates, we use the measured $\text{M}_{\text{V}}$ with $\text{M}_{\star}/L_{\rm V} = 0.3-2$ for the intermediate age and metallicity models found to be most similar to our candidates. We obtain stellar masses of $1.9 \times 10^6 \text{M}_\odot - 1.3 \times 10^7 \text{M}_\odot $ for dwA and $3.1 \times 10^4 \text{M}_\odot - 2.1 \times 10^5 \text{M}_\odot$ for dwB. NGC 628 dwA is therefore similar in mass to the Milky Way's Sculptor and Fornax dwarf spheroidal galaxies, while dwB is most similar in mass to ultrafaint dwarfs like Bootes I, Hercules and Canes Venatici I. Along with NGC dwA, dwB's intermediate age, however, indicates that these candidates are likely not star-forming nor are they ancient re-ionization fossils. This is in sharp contrast with the Local Group ultrafaint dwarfs, which are among the oldest and least chemically-evolved systems known \citep{simon2019}.

We search for signs of on-going star formation in both candidates using ALLWISE \citep{ALLWISE,WISE2010A}, GALEX \citep{GALEX2005} and THINGS H I \citep{walter2008things}. Table \ref{table:candresults} reports the W1, W2 and NUV detections of dwA, along with 5$\sigma$ upper limits for all the non-detections. Using the ALLWISE W1 measurement and $\text{M}_{\star}/\text{L}_{\text{W}1} = 0.6$ for passive, low redshift galaxies and a Chabrier IMF \citep{Kettlety_2017}, we obtain a stellar mass of $3.4 \times 10^6 \text{M}_{\odot}$ for dwA, in line with the mass estimate above. The upper limit on W1 for dwB implies a weak  upper mass limit of $2.6 \times 10^6\text{M}_{\odot}$. Using the z=0 Multiwavelength Galaxy Synthesis (z0MGS) Data Access portal and combining the ALLWISE W1 and W4 and GALEX NUV and based on the results from \citet{Leroy2019}, we obtain upper limits on the star formation rates (SFR) for NGC 628 dwA and dwB of $1.2 \times 10^{-4} \text{M}_{\odot}/\text{yr}$ and $5.4 \times 10^{-5} \text{M}_{\odot}/\text{yr}$, respectively. We additionally find an upper limit on H\textsc{i} gas of $M_{\text{H}\textsc{i}} < 5 \times 10^5 \text{M}_\odot$ for both satellites based on THINGS data.  This certainly means that dwA is gas-poor.  For dwB, the upper limit on $M_{\text{H}\textsc{i}}/M_*$ is about 10, if the stellar mass of dwB is at the low end of our stellar mass estimate.  For the higher, more realistic end of our stellar mass estimate, $M_{\text{H}\textsc{i}}/M_* \lesssim 2$, which is at the very low end of field dwarf mass ratios \citep{papastergis2012}. We consider it likely that dwB, too, is gas-poor.

NGC 628 dwA and dwB therefore have little to no UV flux, low upper limits on SFR, and are gas-poor, which indicate that they are likely not presently star-forming. Their emission in the bluer bands indicates, however, that they were likely star-forming in the recent ($\sim$1 Gyr) past. This is surprising given NGC 628 dwB's luminosity is near the ultrafaint-classicial boundary and indicates that SF may have recently been quenched due to environmental effects, rather than global reionization effects.

Our likely recently quenched candidates stand in contrast to the satellite galaxies of the relatively isolated M94 \citep{smercina2018lonely}, which are all star-forming above this magnitude limit. NGC 628 dwA and dwB are more in line with the mostly quenched satellite populations of the MW \citep[thought to be caused by ram-pressure stripping at infall;][]{slater2014,emerick2016gas,simpson2018quenching}, M31 \citep{McConnachieM31}, M81 \citep{chiboucas2013confirmation}, CenA \citep{crnojevic2019faint}, and M101 \citep{bennet2019m101}. NGC 628 dwA is near the $\text{M}_{\text{V}} <-12.1$ detection limits of the SAGA survey \citep{geha2017saga} which found 26 star-forming satellites out of 27 total around MW-mass hosts. \citet{karunakaran2019} found that, after accounting for completeness, satellites within the virial radius of MW-mass or larger hosts that were brighter than $\text{M}_{\text{V}} \approx -12$ are broadly star-forming and gas-rich, while those fainter than this threshold are broadly quiescent and gas-poor. NGC 628 dwA is likely quenched, and in orbit around a lower-mass host than those in the SAGA sample and in \citet{karunakaran2019}. There are few known dwarfs in close projection to their hosts, as can be seen in \citet{karunakaran2019}, however, dwA is similar in projection to the one quenched SAGA satellite, which could indicate that the relationship between quiescence and radial distance seen for MW-mass hosts could exist for less massive hosts. Therefore, environmental factors beyond simple isolation and group richness, such as strangulation, tidal stripping, and ram pressure stripping could play an important role even for satellite populations of sub MW-mass hosts. Additionally, if the candidates are in fact quenched and at their projected proximity to the host, they would follow the morphology-density relation seen in the Local Group \citep{einasto1974missing,weisz2011,McConnachie12}, which is determined in part by these environmental effects. Recent results from LBT-SONG show a Large Magellanic Cloud-mass host likely quenched star formation in its satellite galaxy via strangulation \citep{garling2019ddo113}, while the MADCASH survey contains a host of similarly low mass that has quenched and tidally disrupted a satellite galaxy \citep{Carlin_2019}. Together with the satellite candidates of NGC 628, these results show environmental effects such as strangulation, and tidal stripping of even low-mass hosts may play an important role in the lives of satellite galaxies.  

\section{NGC 628 satellite counts}
\begin{figure*}[!t]
 \includegraphics[width=\textwidth]{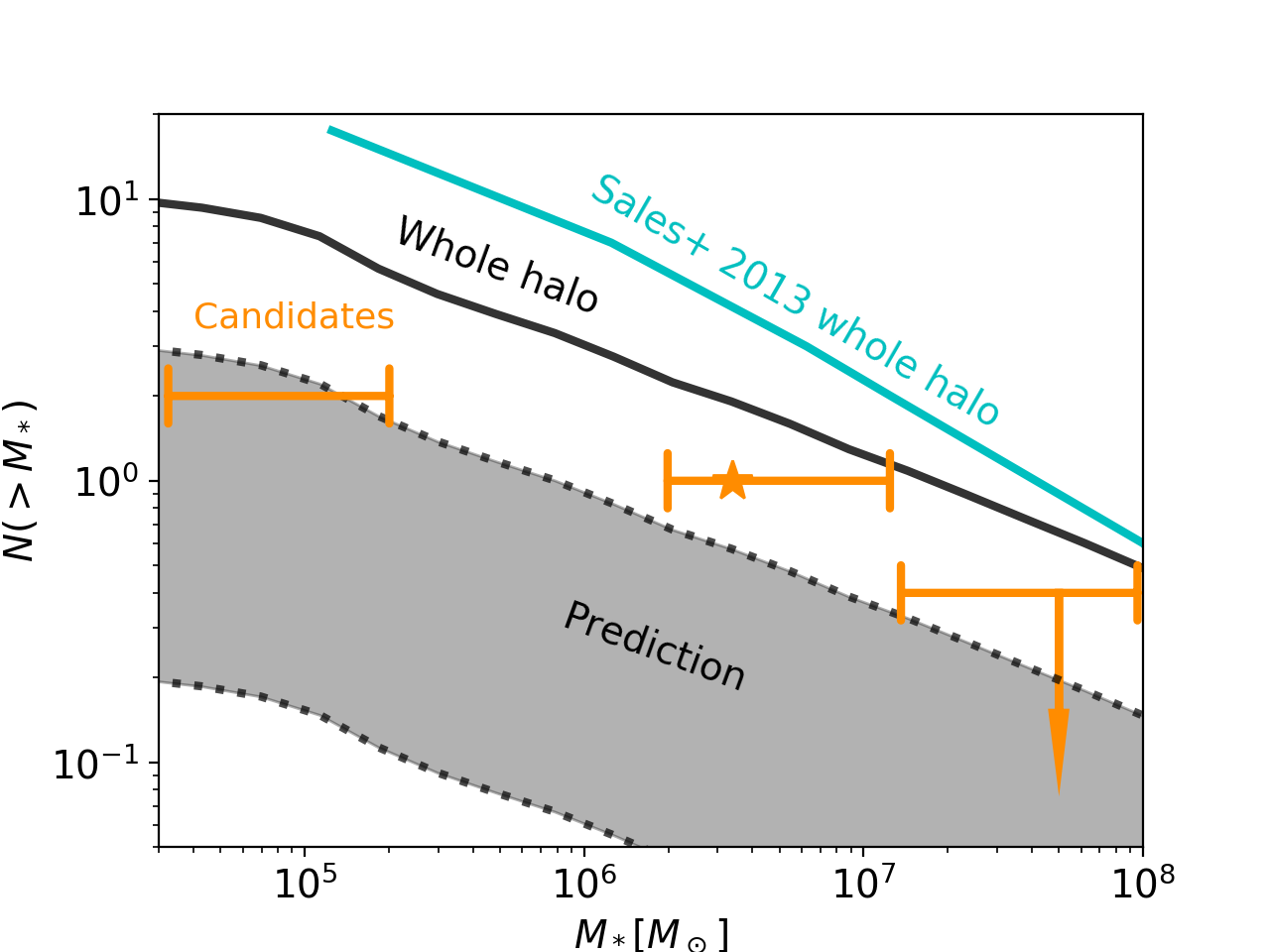}
  \setcounter{figure}{10}
 \caption{The stellar mass function implied by the satellite candidates along with predictions. The cyan line is the whole-halo prediction from \citet{Sales13}. Our whole-halo prediction, described in the text, is indicated with the solid black line. The shaded grey region is our best estimate for the number of detectable satellite galaxies within the LBC footprint, given our measured completeness (Sec.~\ref{sec:completeness}). The width of the region reflects the uncertainty in the radial distribution of satellites, and the galaxy size distribution function. Our satellite candidates are shown in yellow. The yellow star represents the ALLWISE stellar mass estimate for dwA. The error bars represent the minimum and maximum possible $\text{M}_{\star}$ given the extrema of possible $\text{M}_{\star}/\text{L}_{\text{V}}$ ratios, as described in the text. }
 \label{fig:NGCluminosity}
\end{figure*} 

In Fig.~\ref{fig:NGCluminosity}, we compare our inferred satellite stellar mass function to theoretical predictions within the context of the CDM paradigm. The yellow brackets represent the galaxy stellar mass function implied by our satellite candidates and the completeness, folding in the uncertainty in the stellar masses. The lower bound mass estimate for the candidates is obtained by using $\text{M}_{\star}/\text{L}_{\text{V}}$ = 0.35 $\text{M}_\odot/\text{L}_\odot$, the typical mass-to-light ratio for our best-fit 1 Gyr SSP model (Fig.~\ref{fig:color-color}). For the upper stellar mass limit, we use the maximum possible mass-to-light ratio, $\text{M}_{\star}/\text{L}_{\text{V}} = 2 \text{ M}_\odot/\text{L}_\odot$, for an ancient stellar population with a \citet{Chabrier2001} IMF. Based on the SSP fits and the ALLWISE stellar mass estimate for dwA, shown by the yellow star, we expect that the lower bound is closer to truth than the upper bound.

If the candidates are confirmed, then the satellite stellar mass function is consistent with CDM predictions. The grey shaded region bounded by dotted lines in  Fig.~\ref{fig:NGCluminosity} is our prediction for the stellar mass function for the LBC footprint. We use the completeness result from \S \ref{sec:completeness},  that about 50\% of satellites in the LBC footprint in the magnitude range shown in Fig.~\ref{fig:complAdjusted} are detectable by our methods. As the intrinsic surface brightness-luminosity relationships for satellites of hosts less massive than the MW is unknown, we base this relationship on the small MW satellites of which none are young and star forming and so we take the overall completeness of the older models as a better representation of satellite NGC 628 completeness on the whole. The 50\% completeness estimate is an upper limit for the younger models due to the extra masking required for these models, and likely an upper limit for the lower surface brightness models, based on the lower recovery of the faint objects as well as the existence of MW satellites with surface brightness below the threshold we are sensitive to, as can be seen in the bottom panel of  Fig.~\ref{fig:MWDwarfs}. The overall 50\% estimate, combined with the uncertainty on the radial distribution of satellites, leads to our translation of the whole-halo satellite prediction, represented by the solid black line, to a prediction for our measurement. The major source of uncertainty in our theoretical prediction comes from the radial distribution of satellites in the host halo, and in the distribution of stellar populations, V-band magnitude, and surface brightness for fixed stellar mass, which affects how we translate the completeness function from the $\text{M}_\text{V} - \mu_\text{V,eff}$ plane to $\text{M}_{\star}$. 

We consider the radial distribution first. In order to predict the number of galaxies above a fixed $\text{M}_{\star}$ in our footprint, we estimate how many galaxies we expect to find in the whole halo, and then the fraction we expect to find in the LBC footprint. To show the uncertainty in whole-halo predictions for NGC 628, we show a prediction based on \citet{Sales13} and one of our own based on \citet{Kim2018}. The (cyan) \citet{Sales13} prediction is based on the semi-analytic Millennium-II-based catalog by \citet{Guo2011}, and is a good fit to SDSS satellite luminosity functions at low redshift. Our model (black) is based on the relationship between subhalo mass functions and the host's halo mass, and on empirical $\text{M}_{\star}-\text{M}_{\text{halo}}$ relations. We use the \citet{Moster13} $\text{M}_{\star}-\text{M}_{\text{halo}}$ relation with a 0.2 dex scatter to find the probability distribution function of NGC 628's halo mass and to probabilistically populate the subhalos with galaxies. We use the fixed form of the subhalo mass function as a function of host halo mass employed by \citet{Kim2018}, as well as their model for whether halos have luminous baryons or not to account for reionization suppression. Note that we neglect halo-to-halo scatter in the satellite stellar mass function, which is an additional source of uncertainty. The uncertainty in the radial distribution of low-mass satellites is large and the subject of much debate \citep[see][for a discussion]{Hargis2014,Kim2018,Newton2018}. If the surface density of satellites is constant, we expect 4\% of NGC 628's satellites to lie in our footprint. If the satellites follow the host's dark-matter density \citep[approximately in line with ][]{Newton2018}, we expect 50\% of the satellites to lie in our footprint. The Milky Way's classical satellites lie halfway between these two models \citep{Dooley2017a}.

\section{Summary and future survey plans}
The LBT-SONG survey aims to study the satellites of nearby intermediate-mass galaxy systems outside of the Local Group. With a statistical sample of hosts and their satellites, we will be able to more clearly understand the processes that govern star formation and quenching in low mass dwarfs and tease out the various environmental processes at play over a range of host masses. In this paper: 
 
\begin{itemize}
	\item We demonstrated a new method of detecting and quantifying completeness of satellite-galaxies-as-diffuse-sources using {\tt{SExtractor}}, as a function of satellite properties in magnitude, surface brightness, age and metallicity parameter space most commonly seen in the Local Group. 
	
    \item We perform this analysis on NGC 628, an isolated star-forming host with $\sim1/4$ of the stellar mass of the MW and the most distant host in our survey (9.77 Mpc). We present our completeness results for this system. Notably, we are $88\%$ complete to old, $\sim10$ Gyr, system with $\text{M}_{\text{V}}=-9$ and $\mu_{\text{eff}}$ = 25 mag/$\square''$, as well as over $60\%$ complete to old (10 Gyr) systems with $\text{M}_{\text{V}}$ between $-7 \text{ and} -9$ and $\mu_{\text{eff}}$ = 26 and 27 mag/$\square''$.  
    
    \item We found 2 satellite galaxy candidates, NGC 628 dwA and NGC 628 dwB with $\text{M}_{\text{V}} = -12.2$ and $\text{M}_{\text{V}} = -7.7$, respectively. Folding in our two candidates and completeness results, the derived luminosity function for this host is in line with CDM expectations.
    
    \item We estimate the stellar masses of both candidates to be $3.4 \times10^6 \text{M}_{\odot}$ and $3.1 \times 10^4 \text{M}_{\odot} - 2.1 \times 10^5 \text{M}_{\odot}$ for NGC 628 dwA and dwB, respectively. This puts NGC 628 dwA in the classical dwarf galaxy regime, while NGC 628 dwB straddles the boundary between ultrafaint and classical dwarfs in stellar mass. We find no evidence of ongoing star formation with upper limits of $1.2 \times 10^{-4} \text{M}_{\odot}/\text{yr}$ and $5.4 \times 10^{-5} \text{M}_{\odot}/\text{yr}$, respectively. We also place upper limits on the total neutral hydrogen gas mass of $5 \times 10^5 \text{M}_{\odot}$. These data indicate that star formation in the candidates has been recently quenched, on the order of $\sim$1 Gyr ago rather than very early. This would make NGC 628 dwB one of the lowest mass galaxies known that not a reionization fossils, and could indicate that environmental quenching plays a role in modifying satellite populations even for hosts smaller than the MW.
\end{itemize}

Future results from our ground-based survey to identify close satellites of twenty-one Large Magellanic Cloud - to Milky Way-mass galaxies using a combination of resolved star studies on hosts within 3 Mpc to 5  Mpc (Garling et al, in prep), and diffuse galaxy search for hosts within 5 Mpc to 10 Mpc (Davis et al, in prep) will, for the first time, allow us to systematically characterize the close satellite populations of hosts in this mass range outside the Local Group. With completeness quantified as a function of dwarf satellite galaxy properties, we will be able to ascertain the number and type of dwarfs which are present and just as importantly via completeness, those which are not present, around hosts of intermediate mass. Future work using these results along with semi-analytic models will allow us to constrain dwarf galaxy evolution as a function of environment and help to disentangle the myriad astronomical effects shaping the star formation histories of these galaxies, their halos and the true $\text{M}_\star$-$\text{M}_{\text{halo}}$ relation.

\section*{Acknowledgements}
We thank Adam Leroy for helpful discussions. The computations in this paper were run on the CCAPP condos of the Ruby Cluster and the Pitzer Cluster at the Ohio Supercomputer Center \citep{OhioSupercomputerCenter1987}. This material is based upon work supported by the National Science Foundation under Grant No. AST-1615838. AHGP and CTG are also supported by National Science Foundation Grant No. AST-1813628. AMN is supported by a NASA Postdoctoral Program Fellowship. CSK is supported by NSF grants AST-1908952 and AST-1814440. J.P.G. is supported by an NSF Astronomy and Astrophysics Postdoctoral Fellowship under award AST-1801921. Research by DJS is supported by NSF grants AST-1821967, 1821987, 1813708, 1813466, and 1908972. A.S. is supported by an NSF Astronomy and Astrophysics Postdoctoral Fellowship under award AST-1903834.




\bibliographystyle{mnras}
\bibliography{N628}

\begin{thebibliography}{}
\makeatletter
\relax
\def\mn@urlcharsother{\let\do\@makeother \do\$\do\&\do\#\do\^\do\_\do\%\do\~}
\def\mn@doi{\begingroup\mn@urlcharsother \@ifnextchar [ {\mn@doi@}
  {\mn@doi@[]}}
\def\mn@doi@[#1]#2{\def\@tempa{#1}\ifx\@tempa\@empty \href
  {http://dx.doi.org/#2} {doi:#2}\else \href {http://dx.doi.org/#2} {#1}\fi
  \endgroup}
\def\mn@eprint#1#2{\mn@eprint@#1:#2::\@nil}
\def\mn@eprint@arXiv#1{\href {http://arxiv.org/abs/#1} {{\tt arXiv:#1}}}
\def\mn@eprint@dblp#1{\href {http://dblp.uni-trier.de/rec/bibtex/#1.xml}
  {dblp:#1}}
\def\mn@eprint@#1:#2:#3:#4\@nil{\def\@tempa {#1}\def\@tempb {#2}\def\@tempc
  {#3}\ifx \@tempc \@empty \let \@tempc \@tempb \let \@tempb \@tempa \fi \ifx
  \@tempb \@empty \def\@tempb {arXiv}\fi \@ifundefined
  {mn@eprint@\@tempb}{\@tempb:\@tempc}{\expandafter \expandafter \csname
  mn@eprint@\@tempb\endcsname \expandafter{\@tempc}}}

\bibitem[\protect\citeauthoryear{{Adams}, {Kochanek}, {Gerke}  \&
  {Stanek}}{{Adams} et~al.}{2017}]{Adams2017}
{Adams} S.~M.,  {Kochanek} C.~S.,  {Gerke} J.~R.,   {Stanek} K.~Z.,  2017,
  \mn@doi [\mnras] {10.1093/mnras/stx898}, \href
  {http://adsabs.harvard.edu/abs/2017MNRAS.469.1445A} {469, 1445}

\bibitem[\protect\citeauthoryear{Alam et~al.,}{Alam et~al.}{2015}]{Alam2015}
Alam S.,  et~al., 2015, \mn@doi [ApJS] {10.1088/0067-0049/219/1/12}, 219, 12

\bibitem[\protect\citeauthoryear{Aringer, Girardi, Nowotny, Marigo  \&
  Lederer}{Aringer et~al.}{2009}]{Aringer2009}
Aringer B.,  Girardi L.,  Nowotny W.,  Marigo P.,   Lederer M.,  2009,
  Astronomy \& Astrophysics, 503, 913

\bibitem[\protect\citeauthoryear{{Barkana} \& {Loeb}}{{Barkana} \&
  {Loeb}}{1999}]{Barkana1999}
{Barkana} R.,  {Loeb} A.,  1999, \mn@doi [\apj] {10.1086/307724}, \href
  {http://adsabs.harvard.edu/abs/1999ApJ...523...54B} {523, 54}

\bibitem[\protect\citeauthoryear{{Baur}, {Palanque-Delabrouille}, {Y{\`e}che},
  {Magneville}  \& {Viel}}{{Baur} et~al.}{2016}]{Baur_2016}
{Baur} J.,  {Palanque-Delabrouille} N.,  {Y{\`e}che} C.,  {Magneville} C.,
  {Viel} M.,  2016, \mn@doi [\jcap] {10.1088/1475-7516/2016/08/012}, \href
  {https://ui.adsabs.harvard.edu/abs/2016JCAP...08..012B} {2016, 012}

\bibitem[\protect\citeauthoryear{{Behroozi}, {Conroy}  \&
  {Wechsler}}{{Behroozi} et~al.}{2010}]{Behroozi_2010}
{Behroozi} P.~S.,  {Conroy} C.,   {Wechsler} R.~H.,  2010, \mn@doi [\apj]
  {10.1088/0004-637X/717/1/379}, \href
  {https://ui.adsabs.harvard.edu/abs/2010ApJ...717..379B} {717, 379}

\bibitem[\protect\citeauthoryear{{Behroozi}, {Wechsler}  \&
  {Conroy}}{{Behroozi} et~al.}{2013}]{Behroozi13}
{Behroozi} P.~S.,  {Wechsler} R.~H.,   {Conroy} C.,  2013, \mn@doi [\apj]
  {10.1088/0004-637X/770/1/57}, \href
  {http://adsabs.harvard.edu/abs/2013ApJ...770...57B} {770, 57}

\bibitem[\protect\citeauthoryear{{Belokurov} et~al.,}{{Belokurov}
  et~al.}{2007}]{Belokurov07}
{Belokurov} V.,  et~al., 2007, \mn@doi [\apj] {10.1086/509718}, \href
  {http://adsabs.harvard.edu/abs/2007ApJ...654..897B} {654, 897}

\bibitem[\protect\citeauthoryear{Bennet, Sand, Crnojević, Spekkens, Zaritsky
  \& Karunakaran}{Bennet et~al.}{2017}]{Bennet_2017}
Bennet P.,  Sand D.~J.,  Crnojević D.,  Spekkens K.,  Zaritsky D.,
  Karunakaran A.,  2017, \mn@doi [The Astrophysical Journal]
  {10.3847/1538-4357/aa9180}, 850, 109

\bibitem[\protect\citeauthoryear{{Bennet}, {Sand}, {Crnojevi{\'c}}, {Spekkens},
  {Karunakaran}, {Zaritsky}  \& {Mutlu-Pakdil}}{{Bennet}
  et~al.}{2019}]{bennet2019m101}
{Bennet} P.,  {Sand} D.~J.,  {Crnojevi{\'c}} D.,  {Spekkens} K.,  {Karunakaran}
  A.,  {Zaritsky} D.,   {Mutlu-Pakdil} B.,  2019, \mn@doi [\apj]
  {10.3847/1538-4357/ab46ab}, \href
  {https://ui.adsabs.harvard.edu/abs/2019ApJ...885..153B} {885, 153}

\bibitem[\protect\citeauthoryear{{Benson}, {Frenk}, {Lacey}, {Baugh}  \&
  {Cole}}{{Benson} et~al.}{2002}]{Benson02}
{Benson} A.~J.,  {Frenk} C.~S.,  {Lacey} C.~G.,  {Baugh} C.~M.,   {Cole} S.,
  2002, \mn@doi [\mnras] {10.1046/j.1365-8711.2002.05388.x}, \href
  {http://adsabs.harvard.edu/abs/2002MNRAS.333..177B} {333, 177}

\bibitem[\protect\citeauthoryear{Bernstein-Cooper et~al.,}{Bernstein-Cooper
  et~al.}{2014}]{bernstein2014alfalfa}
Bernstein-Cooper E.~Z.,  et~al., 2014, The Astronomical Journal, 148, 35

\bibitem[\protect\citeauthoryear{Bertin}{Bertin}{2006}]{Bertin2006}
Bertin E.,  2006, in {Gabriel} C.,  {Arviset} C.,  {Ponz} D.,   {Enrique} S.,
  eds,  Astronomical Society of the Pacific Conference Series Vol. 351,
  Astronomical Data Analysis Software and Systems XV. p.~112, \url
  {http://adsabs.harvard.edu/abs/2006ASPC..351..112B}

\bibitem[\protect\citeauthoryear{{Bertin} \& {Arnouts}}{{Bertin} \&
  {Arnouts}}{1996}]{Bertin96}
{Bertin} E.,  {Arnouts} S.,  1996, \mn@doi [\aaps] {10.1051/aas:1996164}, \href
  {http://adsabs.harvard.edu/abs/1996A%26AS..117..393B} {117, 393}

\bibitem[\protect\citeauthoryear{Bertin, Mellier, Radovich, Missonnier, Didelon
   \& Morin}{Bertin et~al.}{2002}]{Bertin2002}
Bertin E.,  Mellier Y.,  Radovich M.,  Missonnier G.,  Didelon P.,   Morin B.,
  2002, in {Bohlender} D.~A.,  {Durand} D.,   {Handley} T.~H.,  eds,
  Astronomical Society of the Pacific Conference Series Vol. 281, Astronomical
  Data Analysis Software and Systems XI. p.~228, \url
  {http://adsabs.harvard.edu/abs/2002ASPC..281..228B}

\bibitem[\protect\citeauthoryear{Boettcher et~al.,}{Boettcher
  et~al.}{2013}]{Boettcher2013}
Boettcher E.,  et~al., 2013, \mn@doi [AJ] {10.1088/0004-6256/146/4/94}, 146, 94

\bibitem[\protect\citeauthoryear{{Boylan-Kolchin}, {Bullock}  \&
  {Kaplinghat}}{{Boylan-Kolchin} et~al.}{2011}]{Boylan_Kolchin_2011}
{Boylan-Kolchin} M.,  {Bullock} J.~S.,   {Kaplinghat} M.,  2011, \mn@doi
  [\mnras] {10.1111/j.1745-3933.2011.01074.x}, \href
  {https://ui.adsabs.harvard.edu/abs/2011MNRAS.415L..40B} {415, L40}

\bibitem[\protect\citeauthoryear{Bressan, Marigo, Girardi, Salasnich, Dal~Cero,
  Rubele  \& Nanni}{Bressan et~al.}{2012}]{Bressan2012}
Bressan A.,  Marigo P.,  Girardi L.,  Salasnich B.,  Dal~Cero C.,  Rubele S.,
  Nanni A.,  2012, Monthly Notices of the Royal Astronomical Society, 427, 127

\bibitem[\protect\citeauthoryear{Brooks \& Zolotov}{Brooks \&
  Zolotov}{2014}]{brooksZolotov2014}
Brooks A.~M.,  Zolotov A.,  2014, The Astrophysical Journal, 786, 87

\bibitem[\protect\citeauthoryear{{Brooks}, {Kuhlen}, {Zolotov}  \&
  {Hooper}}{{Brooks} et~al.}{2013}]{Brooks2013}
{Brooks} A.~M.,  {Kuhlen} M.,  {Zolotov} A.,   {Hooper} D.,  2013, \mn@doi
  [\apj] {10.1088/0004-637X/765/1/22}, \href
  {http://adsabs.harvard.edu/abs/2013ApJ...765...22B} {765, 22}

\bibitem[\protect\citeauthoryear{{Brown} et~al.,}{{Brown}
  et~al.}{2014}]{brown2014}
{Brown} T.~M.,  et~al., 2014, \mn@doi [\apj] {10.1088/0004-637X/796/2/91},
  \href {http://adsabs.harvard.edu/abs/2014ApJ...796...91B} {796, 91}

\bibitem[\protect\citeauthoryear{{Bryan} \& {Norman}}{{Bryan} \&
  {Norman}}{1998}]{Bryan98}
{Bryan} G.~L.,  {Norman} M.~L.,  1998, \mn@doi [\apj] {10.1086/305262}, \href
  {http://adsabs.harvard.edu/abs/1998ApJ...495...80B} {495, 80}

\bibitem[\protect\citeauthoryear{{Bullock} \& {Boylan-Kolchin}}{{Bullock} \&
  {Boylan-Kolchin}}{2017}]{Bullock_2017}
{Bullock} J.~S.,  {Boylan-Kolchin} M.,  2017, \mn@doi [\araa]
  {10.1146/annurev-astro-091916-055313}, \href
  {https://ui.adsabs.harvard.edu/abs/2017ARA&A..55..343B} {55, 343}

\bibitem[\protect\citeauthoryear{{Bullock}, {Kravtsov}  \&
  {Weinberg}}{{Bullock} et~al.}{2000}]{Bullock00}
{Bullock} J.~S.,  {Kravtsov} A.~V.,   {Weinberg} D.~H.,  2000, \mn@doi [\apj]
  {10.1086/309279}, \href {http://adsabs.harvard.edu/abs/2000ApJ...539..517B}
  {539, 517}

\bibitem[\protect\citeauthoryear{Cannon et~al.,}{Cannon
  et~al.}{2015}]{cannon2015alfalfa}
Cannon J.~M.,  et~al., 2015, The Astronomical Journal, 149, 72

\bibitem[\protect\citeauthoryear{{Carlin} et~al.,}{{Carlin}
  et~al.}{2016}]{Carlin16}
{Carlin} J.~L.,  et~al., 2016, preprint, \href
  {http://adsabs.harvard.edu/abs/2016arXiv160802591C} {} (\mn@eprint {arXiv}
  {1608.02591})

\bibitem[\protect\citeauthoryear{Carlin et~al.,}{Carlin
  et~al.}{2019}]{Carlin_2019}
Carlin J.~L.,  et~al., 2019, \mn@doi [The Astrophysical Journal]
  {10.3847/1538-4357/ab4c32}, 886, 109

\bibitem[\protect\citeauthoryear{Carlsten, Greco, Beaton  \& Greene}{Carlsten
  et~al.}{2020}]{Carlsten_2020}
Carlsten S.~G.,  Greco J.~P.,  Beaton R.~L.,   Greene J.~E.,  2020, \mn@doi
  [The Astrophysical Journal] {10.3847/1538-4357/ab7758}, 891, 144

\bibitem[\protect\citeauthoryear{Center}{Center}{1987}]{OhioSupercomputerCenter1987}
Center O.~S.,  1987, Ohio Supercomputer Center, \url
  {http://osc.edu/ark:/19495/f5s1ph73}

\bibitem[\protect\citeauthoryear{Chabrier}{Chabrier}{2003}]{Chabrier2001}
Chabrier G.,  2003, Publications of the Astronomical Society of the Pacific,
  115, 763

\bibitem[\protect\citeauthoryear{Chen, Girardi, Bressan, Marigo, Barbieri  \&
  Kong}{Chen et~al.}{2014}]{Chen2014}
Chen Y.,  Girardi L.,  Bressan A.,  Marigo P.,  Barbieri M.,   Kong X.,  2014,
  Monthly Notices of the Royal Astronomical Society, 444, 2525

\bibitem[\protect\citeauthoryear{Chiboucas, Jacobs, Tully  \&
  Karachentsev}{Chiboucas et~al.}{2013}]{chiboucas2013confirmation}
Chiboucas K.,  Jacobs B.~A.,  Tully R.~B.,   Karachentsev I.~D.,  2013, The
  Astronomical Journal, 146, 126

\bibitem[\protect\citeauthoryear{{Conroy}, {Wechsler}  \& {Kravtsov}}{{Conroy}
  et~al.}{2006}]{Conroy_2006}
{Conroy} C.,  {Wechsler} R.~H.,   {Kravtsov} A.~V.,  2006, \mn@doi [\apj]
  {10.1086/503602}, \href
  {https://ui.adsabs.harvard.edu/abs/2006ApJ...647..201C} {647, 201}

\bibitem[\protect\citeauthoryear{{Correa}, {Schaye}, {Wyithe}, {Duffy},
  {Theuns}, {Crain}  \& {Bower}}{{Correa} et~al.}{2018}]{Correa_2017}
{Correa} C.~A.,  {Schaye} J.,  {Wyithe} J. S.~B.,  {Duffy} A.~R.,  {Theuns} T.,
   {Crain} R.~A.,   {Bower} R.~G.,  2018, \mn@doi [\mnras]
  {10.1093/mnras/stx2332}, \href
  {https://ui.adsabs.harvard.edu/abs/2018MNRAS.473..538C} {473, 538}

\bibitem[\protect\citeauthoryear{Crnojevi{\'c} et~al.,}{Crnojevi{\'c}
  et~al.}{2019}]{crnojevic2019faint}
Crnojevi{\'c} D.,  et~al., 2019, The Astrophysical Journal, 872, 80

\bibitem[\protect\citeauthoryear{Cutri et~al.,}{Cutri et~al.}{2013}]{ALLWISE}
Cutri R.,  et~al., 2013, Explanatory Supplement to the AllWISE Data Release
  Products, by RM Cutri et al.

\bibitem[\protect\citeauthoryear{{Dalal} \& {Kochanek}}{{Dalal} \&
  {Kochanek}}{2002}]{Dalal_2002}
{Dalal} N.,  {Kochanek} C.~S.,  2002, \mn@doi [\apj] {10.1086/340303}, \href
  {https://ui.adsabs.harvard.edu/abs/2002ApJ...572...25D} {572, 25}

\bibitem[\protect\citeauthoryear{{Danieli}, {van Dokkum}  \&
  {Conroy}}{{Danieli} et~al.}{2018}]{danieli2018}
{Danieli} S.,  {van Dokkum} P.,   {Conroy} C.,  2018, \mn@doi [\apj]
  {10.3847/1538-4357/aaadfb}, \href
  {http://adsabs.harvard.edu/abs/2018ApJ...856...69D} {856, 69}

\bibitem[\protect\citeauthoryear{{Digby} et~al.,}{{Digby}
  et~al.}{2019}]{digby2019}
{Digby} R.,  et~al., 2019, \mn@doi [\mnras] {10.1093/mnras/stz745}, \href
  {http://adsabs.harvard.edu/abs/2019MNRAS.485.5423D} {485, 5423}

\bibitem[\protect\citeauthoryear{{Dooley}, {Peter}, {Yang}, {Willman},
  {Griffen}  \& {Frebel}}{{Dooley} et~al.}{2017a}]{Dooley2017a}
{Dooley} G.~A.,  {Peter} A. H.~G.,  {Yang} T.,  {Willman} B.,  {Griffen} B.~F.,
    {Frebel} A.,  2017a, \mn@doi [\mnras] {10.1093/mnras/stx1900}, \href
  {https://ui.adsabs.harvard.edu/abs/2017MNRAS.471.4894D} {471, 4894}

\bibitem[\protect\citeauthoryear{{Dooley}, {Peter}, {Carlin}, {Frebel},
  {Bechtol}  \& {Willman}}{{Dooley} et~al.}{2017b}]{Dooley17b}
{Dooley} G.~A.,  {Peter} A. H.~G.,  {Carlin} J.~L.,  {Frebel} A.,  {Bechtol}
  K.,   {Willman} B.,  2017b, \mn@doi [\mnras] {10.1093/mnras/stx2001}, \href
  {https://ui.adsabs.harvard.edu/abs/2017MNRAS.472.1060D} {472, 1060}

\bibitem[\protect\citeauthoryear{Driver et~al.,}{Driver et~al.}{2009}]{Gama}
Driver S.~P.,  et~al., 2009, Astronomy \& Geophysics, 50, 5

\bibitem[\protect\citeauthoryear{{Drlica-Wagner} et~al.,}{{Drlica-Wagner}
  et~al.}{2015}]{DrlicaWagner2015}
{Drlica-Wagner} A.,  et~al., 2015, \mn@doi [\apj]
  {10.1088/0004-637X/813/2/109}, \href
  {http://adsabs.harvard.edu/abs/2015ApJ...813..109D} {813, 109}

\bibitem[\protect\citeauthoryear{Einasto, Saar, Kaasik  \& Chernin}{Einasto
  et~al.}{1974}]{einasto1974missing}
Einasto J.,  Saar E.,  Kaasik A.,   Chernin A.~D.,  1974, Nature, 252, 111

\bibitem[\protect\citeauthoryear{Emerick, Mac~Low, Grcevich  \& Gatto}{Emerick
  et~al.}{2016}]{emerick2016gas}
Emerick A.,  Mac~Low M.-M.,  Grcevich J.,   Gatto A.,  2016, The Astrophysical
  Journal, 826, 148

\bibitem[\protect\citeauthoryear{{Fillingham}, {Cooper}, {Pace},
  {Boylan-Kolchin}, {Bullock}, {Garrison-Kimmel}  \& {Wheeler}}{{Fillingham}
  et~al.}{2016}]{fillingham2016}
{Fillingham} S.~P.,  {Cooper} M.~C.,  {Pace} A.~B.,  {Boylan-Kolchin} M.,
  {Bullock} J.~S.,  {Garrison-Kimmel} S.,   {Wheeler} C.,  2016, \mn@doi
  [\mnras] {10.1093/mnras/stw2131}, \href
  {https://ui.adsabs.harvard.edu/abs/2016MNRAS.463.1916F} {463, 1916}

\bibitem[\protect\citeauthoryear{{Fitts} et~al.,}{{Fitts}
  et~al.}{2017}]{Fitts2017}
{Fitts} A.,  et~al., 2017, \mn@doi [\mnras] {10.1093/mnras/stx1757}, \href
  {https://ui.adsabs.harvard.edu/abs/2017MNRAS.471.3547F} {471, 3547}

\bibitem[\protect\citeauthoryear{Flores \& Primack}{Flores \&
  Primack}{1994}]{Flores_1994}
Flores R.~A.,  Primack J.~R.,  1994, \mn@doi [The Astrophysical Journal]
  {10.1086/187350}, 427, L1

\bibitem[\protect\citeauthoryear{{Gaia Collaboration}, Brown, Vallenari,
  Prusti, de Bruijne, Babusiaux  \& Bailer-Jones}{{Gaia Collaboration}
  et~al.}{2018}]{GaiaCollaboration2018}
{Gaia Collaboration} Brown A. G.~A.,  Vallenari A.,  Prusti T.,  de Bruijne J.
  H.~J.,  Babusiaux C.,   Bailer-Jones C. A.~L.,  2018, \mn@doi [A{\&}A]
  {10.1051/0004-6361/201833051}, 616, A1

\bibitem[\protect\citeauthoryear{{Garling}, {Peter}, {Kochanek}, {Sand}  \&
  {Crnojevi{\'c}}}{{Garling} et~al.}{2020}]{garling2019ddo113}
{Garling} C.~T.,  {Peter} A. H.~G.,  {Kochanek} C.~S.,  {Sand} D.~J.,
  {Crnojevi{\'c}} D.,  2020, \mn@doi [\mnras] {10.1093/mnras/stz3526}, \href
  {https://ui.adsabs.harvard.edu/abs/2020MNRAS.492.1713G} {492, 1713}

\bibitem[\protect\citeauthoryear{{Geha}, {Blanton}, {Yan}  \& {Tinker}}{{Geha}
  et~al.}{2012}]{geha2012}
{Geha} M.,  {Blanton} M.~R.,  {Yan} R.,   {Tinker} J.~L.,  2012, \mn@doi [\apj]
  {10.1088/0004-637X/757/1/85}, \href
  {http://adsabs.harvard.edu/abs/2012ApJ...757...85G} {757, 85}

\bibitem[\protect\citeauthoryear{Geha et~al.,}{Geha
  et~al.}{2017}]{geha2017saga}
Geha M.,  et~al., 2017, The Astrophysical Journal, 847, 4

\bibitem[\protect\citeauthoryear{Gerke, Kochanek  \& Stanek}{Gerke
  et~al.}{2015}]{Gerke2015}
Gerke J.~R.,  Kochanek C.~S.,   Stanek K.~Z.,  2015, \mn@doi [MNRAS]
  {10.1093/mnras/stv776}, 450, 3289

\bibitem[\protect\citeauthoryear{Giallongo et~al.,}{Giallongo
  et~al.}{2008}]{Giallongo2008}
Giallongo E.,  et~al., 2008, Astronomy \& Astrophysics, 482, 349

\bibitem[\protect\citeauthoryear{{Gilman}, {Birrer}, {Nierenberg}, {Treu}, {Du}
   \& {Benson}}{{Gilman} et~al.}{2020}]{gilman2019warm}
{Gilman} D.,  {Birrer} S.,  {Nierenberg} A.,  {Treu} T.,  {Du} X.,   {Benson}
  A.,  2020, \mn@doi [\mnras] {10.1093/mnras/stz3480}, \href
  {https://ui.adsabs.harvard.edu/abs/2020MNRAS.491.6077G} {491, 6077}

\bibitem[\protect\citeauthoryear{{Gnedin}}{{Gnedin}}{2000}]{Gnedin2000}
{Gnedin} N.~Y.,  2000, \mn@doi [\apj] {10.1086/317042}, \href
  {http://adsabs.harvard.edu/abs/2000ApJ...542..535G} {542, 535}

\bibitem[\protect\citeauthoryear{{Greco} et~al.,}{{Greco}
  et~al.}{2018}]{Greco18}
{Greco} J.~P.,  et~al., 2018, \mn@doi [\apj] {10.3847/1538-4357/aab842}, \href
  {http://adsabs.harvard.edu/abs/2018ApJ...857..104G} {857, 104}

\bibitem[\protect\citeauthoryear{{Grossauer} et~al.,}{{Grossauer}
  et~al.}{2015}]{Grossauer15}
{Grossauer} J.,  et~al., 2015, \mn@doi [\apj] {10.1088/0004-637X/807/1/88},
  \href {http://adsabs.harvard.edu/abs/2015ApJ...807...88G} {807, 88}

\bibitem[\protect\citeauthoryear{{Gunn} \& {Gott}}{{Gunn} \&
  {Gott}}{1972}]{gunn1972}
{Gunn} J.~E.,  {Gott} J.~R.~I.,  1972, \mn@doi [\apj] {10.1086/151605}, \href
  {http://adsabs.harvard.edu/abs/1972ApJ...176....1G} {176, 1}

\bibitem[\protect\citeauthoryear{{Guo} et~al.,}{{Guo} et~al.}{2011}]{Guo2011}
{Guo} Q.,  et~al., 2011, \mn@doi [\mnras] {10.1111/j.1365-2966.2010.18114.x},
  \href {https://ui.adsabs.harvard.edu/abs/2011MNRAS.413..101G} {413, 101}

\bibitem[\protect\citeauthoryear{{Hargis}, {Willman}  \& {Peter}}{{Hargis}
  et~al.}{2014}]{Hargis2014}
{Hargis} J.~R.,  {Willman} B.,   {Peter} A.~H.~G.,  2014, \mn@doi [\apjl]
  {10.1088/2041-8205/795/1/L13}, \href
  {http://adsabs.harvard.edu/abs/2014ApJ...795L..13H} {795, L13}

\bibitem[\protect\citeauthoryear{{H{\"a}u{\ss}ler} et~al.,}{{H{\"a}u{\ss}ler}
  et~al.}{2013}]{GALFITM}
{H{\"a}u{\ss}ler} B.,  et~al., 2013, \mn@doi [\mnras] {10.1093/mnras/sts633},
  \href {https://ui.adsabs.harvard.edu/abs/2013MNRAS.430..330H} {430, 330}

\bibitem[\protect\citeauthoryear{{Homma} et~al.,}{{Homma}
  et~al.}{2018}]{Homma2018}
{Homma} D.,  et~al., 2018, \mn@doi [\pasj] {10.1093/pasj/psx050}, \href
  {http://adsabs.harvard.edu/abs/2018PASJ...70S..18H} {70, S18}

\bibitem[\protect\citeauthoryear{{Hu}, {Barkana}  \& {Gruzinov}}{{Hu}
  et~al.}{2000}]{Hu2000}
{Hu} W.,  {Barkana} R.,   {Gruzinov} A.,  2000, \mn@doi [Physical Review
  Letters] {10.1103/PhysRevLett.85.1158}, \href
  {http://adsabs.harvard.edu/abs/2000PhRvL..85.1158H} {85, 1158}

\bibitem[\protect\citeauthoryear{{Hui}, {Ostriker}, {Tremaine}  \&
  {Witten}}{{Hui} et~al.}{2017}]{Hui2017}
{Hui} L.,  {Ostriker} J.~P.,  {Tremaine} S.,   {Witten} E.,  2017, \mn@doi
  [\prd] {10.1103/PhysRevD.95.043541}, \href
  {http://adsabs.harvard.edu/abs/2017PhRvD..95d3541H} {95, 043541}

\bibitem[\protect\citeauthoryear{{Ir{\v{s}}i{\v{c}}}
  et~al.,}{{Ir{\v{s}}i{\v{c}}} et~al.}{2017}]{Irsic2017}
{Ir{\v{s}}i{\v{c}}} V.,  et~al., 2017, \mn@doi [\prd]
  {10.1103/PhysRevD.96.023522}, \href
  {https://ui.adsabs.harvard.edu/abs/2017PhRvD..96b3522I} {96, 023522}

\bibitem[\protect\citeauthoryear{{Jethwa}, {Erkal}  \& {Belokurov}}{{Jethwa}
  et~al.}{2018}]{Jethwa_2017}
{Jethwa} P.,  {Erkal} D.,   {Belokurov} V.,  2018, \mn@doi [\mnras]
  {10.1093/mnras/stx2330}, \href
  {https://ui.adsabs.harvard.edu/abs/2018MNRAS.473.2060J} {473, 2060}

\bibitem[\protect\citeauthoryear{Jordi, Grebel  \& Ammon}{Jordi
  et~al.}{2006}]{Jordi2006}
Jordi K.,  Grebel E.~K.,   Ammon K.,  2006, \mn@doi [A{\&}A]
  {10.1051/0004-6361:20066082}, 460, 339

\bibitem[\protect\citeauthoryear{Karachentsev, Makarov  \&
  Kaisina}{Karachentsev et~al.}{2013}]{Karachentsev2013updated}
Karachentsev I.~D.,  Makarov D.~I.,   Kaisina E.~I.,  2013, The Astronomical
  Journal, 145, 101

\bibitem[\protect\citeauthoryear{{Karunakaran}, {Spekkens}, {Bennet}, {Sand},
  {Crnojevi{\'c}}  \& {Zaritsky}}{{Karunakaran} et~al.}{2020}]{karunakaran2019}
{Karunakaran} A.,  {Spekkens} K.,  {Bennet} P.,  {Sand} D.~J.,  {Crnojevi{\'c}}
  D.,   {Zaritsky} D.,  2020, \mn@doi [\aj] {10.3847/1538-3881/ab5af1}, \href
  {https://ui.adsabs.harvard.edu/abs/2020AJ....159...37K} {159, 37}

\bibitem[\protect\citeauthoryear{{Kauffmann}, {White}  \&
  {Guiderdoni}}{{Kauffmann} et~al.}{1993}]{kauffmann93}
{Kauffmann} G.,  {White} S.~D.~M.,   {Guiderdoni} B.,  1993, \mn@doi [\mnras]
  {10.1093/mnras/264.1.201}, \href
  {https://ui.adsabs.harvard.edu/abs/1993MNRAS.264..201K} {264, 201}

\bibitem[\protect\citeauthoryear{{Kettlety} et~al.,}{{Kettlety}
  et~al.}{2018}]{Kettlety_2017}
{Kettlety} T.,  et~al., 2018, \mn@doi [\mnras] {10.1093/mnras/stx2379}, \href
  {https://ui.adsabs.harvard.edu/abs/2018MNRAS.473..776K} {473, 776}

\bibitem[\protect\citeauthoryear{Kim, Kim, Hwang, Lee, Chun  \& Ann}{Kim
  et~al.}{2011}]{Kim2011}
Kim E.,  Kim M.,  Hwang N.,  Lee M.~G.,  Chun M.-Y.,   Ann H.~B.,  2011,
  \mn@doi [Monthly Notices of the Royal Astronomical Society]
  {10.1111/j.1365-2966.2010.18022.x}, 412, 1881–1894

\bibitem[\protect\citeauthoryear{{Kim}, {Jerjen}, {Mackey}, {Da Costa}  \&
  {Milone}}{{Kim} et~al.}{2015}]{Kim2015}
{Kim} D.,  {Jerjen} H.,  {Mackey} D.,  {Da Costa} G.~S.,   {Milone} A.~P.,
  2015, \mn@doi [\apjl] {10.1088/2041-8205/804/2/L44}, \href
  {http://adsabs.harvard.edu/abs/2015ApJ...804L..44K} {804, L44}

\bibitem[\protect\citeauthoryear{{Kim}, {Peter}  \& {Hargis}}{{Kim}
  et~al.}{2018}]{Kim2018}
{Kim} S.~Y.,  {Peter} A.~H.~G.,   {Hargis} J.~R.,  2018, \mn@doi [Physical
  Review Letters] {10.1103/PhysRevLett.121.211302}, \href
  {http://adsabs.harvard.edu/abs/2018PhRvL.121u1302K} {121, 211302}

\bibitem[\protect\citeauthoryear{{Kirby}, {Cohen}, {Guhathakurta}, {Cheng},
  {Bullock}  \& {Gallazzi}}{{Kirby} et~al.}{2013}]{Kirby13}
{Kirby} E.~N.,  {Cohen} J.~G.,  {Guhathakurta} P.,  {Cheng} L.,  {Bullock}
  J.~S.,   {Gallazzi} A.,  2013, \mn@doi [\apj] {10.1088/0004-637X/779/2/102},
  \href {http://adsabs.harvard.edu/abs/2013ApJ...779..102K} {779, 102}

\bibitem[\protect\citeauthoryear{{Klypin}, {Kravtsov}, {Valenzuela}  \&
  {Prada}}{{Klypin} et~al.}{1999a}]{Klypin99}
{Klypin} A.,  {Kravtsov} A.~V.,  {Valenzuela} O.,   {Prada} F.,  1999a, \mn@doi
  [\apj] {10.1086/307643}, \href
  {http://adsabs.harvard.edu/abs/1999ApJ...522...82K} {522, 82}

\bibitem[\protect\citeauthoryear{{Klypin}, {Kravtsov}, {Valenzuela}  \&
  {Prada}}{{Klypin} et~al.}{1999b}]{KlypinMissingSats}
{Klypin} A.,  {Kravtsov} A.~V.,  {Valenzuela} O.,   {Prada} F.,  1999b, \mn@doi
  [\apj] {10.1086/307643}, \href
  {https://ui.adsabs.harvard.edu/abs/1999ApJ...522...82K} {522, 82}

\bibitem[\protect\citeauthoryear{{Kochanek}, {Beacom}, {Kistler}, {Prieto},
  {Stanek}, {Thompson}  \& {Y{\"u}ksel}}{{Kochanek} et~al.}{2008}]{Kochanek08}
{Kochanek} C.~S.,  {Beacom} J.~F.,  {Kistler} M.~D.,  {Prieto} J.~L.,  {Stanek}
  K.~Z.,  {Thompson} T.~A.,   {Y{\"u}ksel} H.,  2008, \mn@doi [\apj]
  {10.1086/590053}, \href {http://adsabs.harvard.edu/abs/2008ApJ...684.1336K}
  {684, 1336}

\bibitem[\protect\citeauthoryear{{Koposov} et~al.,}{{Koposov}
  et~al.}{2008}]{Koposov2008}
{Koposov} S.,  et~al., 2008, \mn@doi [\apj] {10.1086/589911}, \href
  {http://adsabs.harvard.edu/abs/2008ApJ...686..279K} {686, 279}

\bibitem[\protect\citeauthoryear{{Laevens} et~al.,}{{Laevens}
  et~al.}{2015}]{Laevens2015}
{Laevens} B.~P.~M.,  et~al., 2015, \mn@doi [\apj] {10.1088/0004-637X/813/1/44},
  \href {http://adsabs.harvard.edu/abs/2015ApJ...813...44L} {813, 44}

\bibitem[\protect\citeauthoryear{Lang, Hogg, Mierle, Blanton  \& Roweis}{Lang
  et~al.}{2010}]{Lang2010}
Lang D.,  Hogg D.~W.,  Mierle K.,  Blanton M.,   Roweis S.,  2010, \mn@doi [AJ]
  {10.1088/0004-6256/139/5/1782}, 139, 1782

\bibitem[\protect\citeauthoryear{{Larson}, {Tinsley}  \& {Caldwell}}{{Larson}
  et~al.}{1980}]{larson1980}
{Larson} R.~B.,  {Tinsley} B.~M.,   {Caldwell} C.~N.,  1980, \mn@doi [ApJ]
  {10.1086/157917}, \href {http://adsabs.harvard.edu/abs/1980ApJ...237..692L}
  {237, 692}

\bibitem[\protect\citeauthoryear{Leroy, Walter, Brinks, Bigiel, De~Blok, Madore
   \& Thornley}{Leroy et~al.}{2008}]{Leroy2008THINGS}
Leroy A.~K.,  Walter F.,  Brinks E.,  Bigiel F.,  De~Blok W.,  Madore B.,
  Thornley M.,  2008, The Astronomical Journal, 136, 2782

\bibitem[\protect\citeauthoryear{{Leroy} et~al.,}{{Leroy}
  et~al.}{2019}]{Leroy2019}
{Leroy} A.~K.,  et~al., 2019, \mn@doi [\apjs] {10.3847/1538-4365/ab3925}, \href
  {https://ui.adsabs.harvard.edu/abs/2019ApJS..244...24L} {244, 24}

\bibitem[\protect\citeauthoryear{{Loveday} et~al.,}{{Loveday}
  et~al.}{2015}]{Loveday2015}
{Loveday} J.,  et~al., 2015, \mn@doi [\mnras] {10.1093/mnras/stv1013}, \href
  {http://adsabs.harvard.edu/abs/2015MNRAS.451.1540L} {451, 1540}

\bibitem[\protect\citeauthoryear{{Lovell} et~al.,}{{Lovell}
  et~al.}{2017}]{Lovell2017}
{Lovell} M.~R.,  et~al., 2017, \mn@doi [\mnras] {10.1093/mnras/stx654}, \href
  {http://adsabs.harvard.edu/abs/2017MNRAS.468.4285L} {468, 4285}

\bibitem[\protect\citeauthoryear{{Marigo} et~al.,}{{Marigo}
  et~al.}{2017}]{Marigo17}
{Marigo} P.,  et~al., 2017, \mn@doi [\apj] {10.3847/1538-4357/835/1/77}, \href
  {http://adsabs.harvard.edu/abs/2017ApJ...835...77M} {835, 77}

\bibitem[\protect\citeauthoryear{{Martin} et~al.,}{{Martin}
  et~al.}{2005}]{GALEX2005}
{Martin} D.~C.,  et~al., 2005, \mn@doi [\apjl] {10.1086/426387}, \href
  {https://ui.adsabs.harvard.edu/abs/2005ApJ...619L...1M} {619, L1}

\bibitem[\protect\citeauthoryear{{Mayer}, {Mastropietro}, {Wadsley}, {Stadel}
  \& {Moore}}{{Mayer} et~al.}{2006}]{mayer2006}
{Mayer} L.,  {Mastropietro} C.,  {Wadsley} J.,  {Stadel} J.,   {Moore} B.,
  2006, \mn@doi [\mnras] {10.1111/j.1365-2966.2006.10403.x}, \href
  {http://adsabs.harvard.edu/abs/2006MNRAS.369.1021M} {369, 1021}

\bibitem[\protect\citeauthoryear{{McConnachie}}{{McConnachie}}{2012}]{McConnachie12}
{McConnachie} A.~W.,  2012, \mn@doi [\aj] {10.1088/0004-6256/144/1/4}, \href
  {http://adsabs.harvard.edu/abs/2012AJ....144....4M} {144, 4}

\bibitem[\protect\citeauthoryear{McConnachie \& Irwin}{McConnachie \&
  Irwin}{2006}]{McConnachieM31}
McConnachie A.~W.,  Irwin M.~J.,  2006, \mn@doi [Monthly Notices of the Royal
  Astronomical Society] {10.1111/j.1365-2966.2005.09771.x}, 365, 902

\bibitem[\protect\citeauthoryear{McQuinn, Skillman, Dolphin, Berg  \&
  Kennicutt}{McQuinn et~al.}{2017}]{Mcquinn2017Distance}
McQuinn K.~B.,  Skillman E.~D.,  Dolphin A.~E.,  Berg D.,   Kennicutt R.,
  2017, The Astronomical Journal, 154, 51

\bibitem[\protect\citeauthoryear{Merritt, Van~Dokkum  \& Abraham}{Merritt
  et~al.}{2014}]{merritt2014discovery}
Merritt A.,  Van~Dokkum P.,   Abraham R.,  2014, The Astrophysical Journal
  Letters, 787, L37

\bibitem[\protect\citeauthoryear{{Moore}}{{Moore}}{1994}]{Moore1994}
{Moore} B.,  1994, \mn@doi [\nat] {10.1038/370629a0}, \href
  {https://ui.adsabs.harvard.edu/abs/1994Natur.370..629M} {370, 629}

\bibitem[\protect\citeauthoryear{{Moore}, {Quinn}, {Governato}, {Stadel}  \&
  {Lake}}{{Moore} et~al.}{1999a}]{Moore_1999}
{Moore} B.,  {Quinn} T.,  {Governato} F.,  {Stadel} J.,   {Lake} G.,  1999a,
  \mn@doi [\mnras] {10.1046/j.1365-8711.1999.03039.x}, \href
  {https://ui.adsabs.harvard.edu/abs/1999MNRAS.310.1147M} {310, 1147}

\bibitem[\protect\citeauthoryear{{Moore}, {Ghigna}, {Governato}, {Lake},
  {Quinn}, {Stadel}  \& {Tozzi}}{{Moore} et~al.}{1999b}]{Moore99}
{Moore} B.,  {Ghigna} S.,  {Governato} F.,  {Lake} G.,  {Quinn} T.,  {Stadel}
  J.,   {Tozzi} P.,  1999b, \mn@doi [\apjl] {10.1086/312287}, \href
  {http://adsabs.harvard.edu/abs/1999ApJ...524L..19M} {524, L19}

\bibitem[\protect\citeauthoryear{{Moster}, {Naab}  \& {White}}{{Moster}
  et~al.}{2013}]{Moster13}
{Moster} B.~P.,  {Naab} T.,   {White} S.~D.~M.,  2013, \mn@doi [\mnras]
  {10.1093/mnras/sts261}, \href
  {http://adsabs.harvard.edu/abs/2013MNRAS.428.3121M} {428, 3121}

\bibitem[\protect\citeauthoryear{{Mu{\~n}oz}, {C{\^o}t{\'e}}, {Santana},
  {Geha}, {Simon}, {Oyarz{\'u}n}, {Stetson}  \& {Djorgovski}}{{Mu{\~n}oz}
  et~al.}{2018}]{Munoz18}
{Mu{\~n}oz} R.~R.,  {C{\^o}t{\'e}} P.,  {Santana} F.~A.,  {Geha} M.,  {Simon}
  J.~D.,  {Oyarz{\'u}n} G.~A.,  {Stetson} P.~B.,   {Djorgovski} S.~G.,  2018,
  \mn@doi [\apj] {10.3847/1538-4357/aac16b}, \href
  {https://ui.adsabs.harvard.edu/abs/2018ApJ...860...66M} {860, 66}

\bibitem[\protect\citeauthoryear{M{\"u}ller, Scalera, Binggeli  \&
  Jerjen}{M{\"u}ller et~al.}{2017}]{muller2017m}
M{\"u}ller O.,  Scalera R.,  Binggeli B.,   Jerjen H.,  2017, Astronomy \&
  Astrophysics, 602, A119

\bibitem[\protect\citeauthoryear{{M{\"u}ller}, {Jerjen}  \&
  {Binggeli}}{{M{\"u}ller} et~al.}{2018}]{Muller2018a}
{M{\"u}ller} O.,  {Jerjen} H.,   {Binggeli} B.,  2018, \mn@doi [\aap]
  {10.1051/0004-6361/201832897}, \href
  {https://ui.adsabs.harvard.edu/abs/2018A&A...615A.105M} {615, A105}

\bibitem[\protect\citeauthoryear{{Munshi} et~al.,}{{Munshi}
  et~al.}{2013}]{Munshi13}
{Munshi} F.,  et~al., 2013, \mn@doi [\apj] {10.1088/0004-637X/766/1/56}, \href
  {http://adsabs.harvard.edu/abs/2013ApJ...766...56M} {766, 56}

\bibitem[\protect\citeauthoryear{{Munshi}, {Brooks}, {Christensen},
  {Applebaum}, {Holley-Bockelmann}, {Quinn}  \& {Wadsley}}{{Munshi}
  et~al.}{2019}]{munshi2019}
{Munshi} F.,  {Brooks} A.~M.,  {Christensen} C.,  {Applebaum} E.,
  {Holley-Bockelmann} K.,  {Quinn} T.~R.,   {Wadsley} J.,  2019, \mn@doi [\apj]
  {10.3847/1538-4357/ab0085}, \href
  {https://ui.adsabs.harvard.edu/abs/2019ApJ...874...40M} {874, 40}

\bibitem[\protect\citeauthoryear{{Nadler}, {Mao}, {Green}  \&
  {Wechsler}}{{Nadler} et~al.}{2019a}]{Nadler2019}
{Nadler} E.~O.,  {Mao} Y.-Y.,  {Green} G.~M.,   {Wechsler} R.~H.,  2019a,
  \mn@doi [\apj] {10.3847/1538-4357/ab040e}, \href
  {http://adsabs.harvard.edu/abs/2019ApJ...873...34N} {873, 34}

\bibitem[\protect\citeauthoryear{Nadler, Mao, Green  \& Wechsler}{Nadler
  et~al.}{2019b}]{Nadler_2019}
Nadler E.~O.,  Mao Y.-Y.,  Green G.~M.,   Wechsler R.~H.,  2019b, \mn@doi [The
  Astrophysical Journal] {10.3847/1538-4357/ab040e}, 873, 34

\bibitem[\protect\citeauthoryear{{Navarro}, {Frenk}  \& {White}}{{Navarro}
  et~al.}{1997}]{NFW1997}
{Navarro} J.~F.,  {Frenk} C.~S.,   {White} S. D.~M.,  1997, \mn@doi [\apj]
  {10.1086/304888}, \href
  {https://ui.adsabs.harvard.edu/abs/1997ApJ...490..493N} {490, 493}

\bibitem[\protect\citeauthoryear{{Newton}, {Cautun}, {Jenkins}, {Frenk}  \&
  {Helly}}{{Newton} et~al.}{2018}]{Newton2018}
{Newton} O.,  {Cautun} M.,  {Jenkins} A.,  {Frenk} C.~S.,   {Helly} J.~C.,
  2018, \mn@doi [\mnras] {10.1093/mnras/sty1085}, \href
  {http://adsabs.harvard.edu/abs/2018MNRAS.479.2853N} {479, 2853}

\bibitem[\protect\citeauthoryear{{Nichols} \& {Bland-Hawthorn}}{{Nichols} \&
  {Bland-Hawthorn}}{2011}]{nichols2011}
{Nichols} M.,  {Bland-Hawthorn} J.,  2011, \mn@doi [\apj]
  {10.1088/0004-637X/732/1/17}, \href
  {http://adsabs.harvard.edu/abs/2011ApJ...732...17N} {732, 17}

\bibitem[\protect\citeauthoryear{{Nierenberg} et~al.,}{{Nierenberg}
  et~al.}{2020}]{nierenberg2019double}
{Nierenberg} A.~M.,  et~al., 2020, \mn@doi [\mnras] {10.1093/mnras/stz3588},
  \href {https://ui.adsabs.harvard.edu/abs/2020MNRAS.492.5314N} {492, 5314}

\bibitem[\protect\citeauthoryear{Papastergis, Cattaneo, Huang, Giovanelli  \&
  Haynes}{Papastergis et~al.}{2012}]{papastergis2012}
Papastergis E.,  Cattaneo A.,  Huang S.,  Giovanelli R.,   Haynes M.~P.,  2012,
  The Astrophysical Journal, 759, 138

\bibitem[\protect\citeauthoryear{Park, Moon, Zaritsky, Pak, Lee, Kim, Kim  \&
  Cha}{Park et~al.}{2017}]{Park2017}
Park H.~S.,  Moon D.-S.,  Zaritsky D.,  Pak M.,  Lee J.-J.,  Kim S.~C.,  Kim
  D.-J.,   Cha S.-M.,  2017, \mn@doi [The Astrophysical Journal]
  {10.3847/1538-4357/aa88ab}, 848, 19

\bibitem[\protect\citeauthoryear{Peng, Ho, Impey  \& Rix}{Peng
  et~al.}{2002}]{GALFIT}
Peng C.~Y.,  Ho L.~C.,  Impey C.~D.,   Rix H.-W.,  2002, The Astronomical
  Journal, 124, 266

\bibitem[\protect\citeauthoryear{{Plummer}}{{Plummer}}{1911}]{Plummer1911}
{Plummer} H.~C.,  1911, \mn@doi [\mnras] {10.1093/mnras/71.5.460}, \href
  {https://ui.adsabs.harvard.edu/abs/1911MNRAS..71..460P} {71, 460}

\bibitem[\protect\citeauthoryear{{Pontzen} \& {Governato}}{{Pontzen} \&
  {Governato}}{2012}]{pontzen2012}
{Pontzen} A.,  {Governato} F.,  2012, \mn@doi [\mnras]
  {10.1111/j.1365-2966.2012.20571.x}, \href
  {https://ui.adsabs.harvard.edu/abs/2012MNRAS.421.3464P} {421, 3464}

\bibitem[\protect\citeauthoryear{{Read} \& {Gilmore}}{{Read} \&
  {Gilmore}}{2005}]{Read2005}
{Read} J.~I.,  {Gilmore} G.,  2005, \mn@doi [\mnras]
  {10.1111/j.1365-2966.2004.08424.x}, \href
  {http://adsabs.harvard.edu/abs/2005MNRAS.356..107R} {356, 107}

\bibitem[\protect\citeauthoryear{{Read}, {Walker}  \& {Steger}}{{Read}
  et~al.}{2019}]{Read2019}
{Read} J.~I.,  {Walker} M.~G.,   {Steger} P.,  2019, \mn@doi [\mnras]
  {10.1093/mnras/sty3404}, \href
  {https://ui.adsabs.harvard.edu/abs/2019MNRAS.484.1401R} {484, 1401}

\bibitem[\protect\citeauthoryear{Reddick, Wechsler, Tinker  \&
  Behroozi}{Reddick et~al.}{2013}]{Reddick_2013}
Reddick R.~M.,  Wechsler R.~H.,  Tinker J.~L.,   Behroozi P.~S.,  2013, \mn@doi
  [The Astrophysical Journal] {10.1088/0004-637x/771/1/30}, 771, 30

\bibitem[\protect\citeauthoryear{{Rodriguez Wimberly}, {Cooper}, {Fillingham},
  {Boylan-Kolchin}, {Bullock}  \& {Garrison-Kimmel}}{{Rodriguez Wimberly}
  et~al.}{2019}]{rodriguezwimberly2019}
{Rodriguez Wimberly} M.~K.,  {Cooper} M.~C.,  {Fillingham} S.~P.,
  {Boylan-Kolchin} M.,  {Bullock} J.~S.,   {Garrison-Kimmel} S.,  2019, \mn@doi
  [\mnras] {10.1093/mnras/sty3357}, \href
  {http://adsabs.harvard.edu/abs/2019MNRAS.483.4031R} {483, 4031}

\bibitem[\protect\citeauthoryear{{Sales}, {Wang}, {White}  \&
  {Navarro}}{{Sales} et~al.}{2013}]{Sales13}
{Sales} L.~V.,  {Wang} W.,  {White} S.~D.~M.,   {Navarro} J.~F.,  2013, \mn@doi
  [\mnras] {10.1093/mnras/sts054}, \href
  {http://adsabs.harvard.edu/abs/2013MNRAS.428..573S} {428, 573}

\bibitem[\protect\citeauthoryear{Sand et~al.,}{Sand
  et~al.}{2015}]{sand2015comprehensive}
Sand D.,  et~al., 2015, The Astrophysical Journal, 806, 95

\bibitem[\protect\citeauthoryear{Sawala et~al.,}{Sawala
  et~al.}{2015}]{sawalaBentByBaryons2015}
Sawala T.,  et~al., 2015, \mn@doi [Monthly Notices of the Royal Astronomical
  Society] {10.1093/mnras/stu2753}, 448, 2941

\bibitem[\protect\citeauthoryear{{Schaye} et~al.,}{{Schaye}
  et~al.}{2015}]{schaye2015}
{Schaye} J.,  et~al., 2015, \mn@doi [\mnras] {10.1093/mnras/stu2058}, \href
  {https://ui.adsabs.harvard.edu/abs/2015MNRAS.446..521S} {446, 521}

\bibitem[\protect\citeauthoryear{{Schlafly} \& {Finkbeiner}}{{Schlafly} \&
  {Finkbeiner}}{2011}]{Schlafly11}
{Schlafly} E.~F.,  {Finkbeiner} D.~P.,  2011, \mn@doi [\apj]
  {10.1088/0004-637X/737/2/103}, \href
  {http://adsabs.harvard.edu/abs/2011ApJ...737..103S} {737, 103}

\bibitem[\protect\citeauthoryear{Schlegel, Finkbeiner  \& Davis}{Schlegel
  et~al.}{1998}]{Schlegel1998}
Schlegel D.~J.,  Finkbeiner D.~P.,   Davis M.,  1998, \mn@doi [ApJ]
  {10.1086/305772}, 500, 525

\bibitem[\protect\citeauthoryear{{Simon}}{{Simon}}{2019}]{simon2019}
{Simon} J.~D.,  2019, arXiv e-prints, \href
  {http://adsabs.harvard.edu/abs/2019arXiv190105465S} {}

\bibitem[\protect\citeauthoryear{{Simpson}, {Bryan}, {Hummels}  \&
  {Ostriker}}{{Simpson} et~al.}{2015}]{Simpson2015}
{Simpson} C.~M.,  {Bryan} G.~L.,  {Hummels} C.,   {Ostriker} J.~P.,  2015,
  \mn@doi [\apj] {10.1088/0004-637X/809/1/69}, \href
  {https://ui.adsabs.harvard.edu/abs/2015ApJ...809...69S} {809, 69}

\bibitem[\protect\citeauthoryear{Simpson, Grand, G{\'o}mez, Marinacci, Pakmor,
  Springel, Campbell  \& Frenk}{Simpson et~al.}{2018}]{simpson2018quenching}
Simpson C.~M.,  Grand R.~J.,  G{\'o}mez F.~A.,  Marinacci F.,  Pakmor R.,
  Springel V.,  Campbell D.~J.,   Frenk C.~S.,  2018, Monthly Notices of the
  Royal Astronomical Society, 478, 548

\bibitem[\protect\citeauthoryear{{Slater} \& {Bell}}{{Slater} \&
  {Bell}}{2014}]{slater2014}
{Slater} C.~T.,  {Bell} E.~F.,  2014, \mn@doi [\apj]
  {10.1088/0004-637X/792/2/141}, \href
  {http://adsabs.harvard.edu/abs/2014ApJ...792..141S} {792, 141}

\bibitem[\protect\citeauthoryear{{Smercina}, {Bell}, {Price}, {D'Souza},
  {Slater}, {Bailin}, {Monachesi}  \& {Nidever}}{{Smercina}
  et~al.}{2018}]{smercina2018lonely}
{Smercina} A.,  {Bell} E.~F.,  {Price} P.~A.,  {D'Souza} R.,  {Slater} C.~T.,
  {Bailin} J.,  {Monachesi} A.,   {Nidever} D.,  2018, \mn@doi [\apj]
  {10.3847/1538-4357/aad2d6}, \href
  {https://ui.adsabs.harvard.edu/abs/2018ApJ...863..152S} {863, 152}

\bibitem[\protect\citeauthoryear{Stetson}{Stetson}{1987}]{Stetson1987}
Stetson P.~B.,  1987, \mn@doi [PASP] {10.1086/131977}, 99, 191

\bibitem[\protect\citeauthoryear{Stetson}{Stetson}{1994}]{Stetson1994}
Stetson P.,  1994, PASP, 106, 250

\bibitem[\protect\citeauthoryear{{Strigari}, {Bullock}  \&
  {Kaplinghat}}{{Strigari} et~al.}{2007}]{Strigari_2007}
{Strigari} L.~E.,  {Bullock} J.~S.,   {Kaplinghat} M.,  2007, \mn@doi [\apjl]
  {10.1086/512976}, \href
  {https://ui.adsabs.harvard.edu/abs/2007ApJ...657L...1S} {657, L1}

\bibitem[\protect\citeauthoryear{Tasitsiomi, Kravtsov, Wechsler  \&
  Primack}{Tasitsiomi et~al.}{2004}]{Tasitsiomi_2004}
Tasitsiomi A.,  Kravtsov A.~V.,  Wechsler R.~H.,   Primack J.~R.,  2004,
  \mn@doi [The Astrophysical Journal] {10.1086/423784}, 614, 533–546

\bibitem[\protect\citeauthoryear{{Tollerud}, {Bullock}, {Strigari}  \&
  {Willman}}{{Tollerud} et~al.}{2008}]{Tollerud2008}
{Tollerud} E.~J.,  {Bullock} J.~S.,  {Strigari} L.~E.,   {Willman} B.,  2008,
  \mn@doi [\apj] {10.1086/592102}, \href
  {http://adsabs.harvard.edu/abs/2008ApJ...688..277T} {688, 277}

\bibitem[\protect\citeauthoryear{{Tollerud}, {Bullock}, {Graves}  \&
  {Wolf}}{{Tollerud} et~al.}{2011}]{Tollerud2011}
{Tollerud} E.~J.,  {Bullock} J.~S.,  {Graves} G.~J.,   {Wolf} J.,  2011,
  \mn@doi [\apj] {10.1088/0004-637X/726/2/108}, \href
  {https://ui.adsabs.harvard.edu/abs/2011ApJ...726..108T} {726, 108}

\bibitem[\protect\citeauthoryear{Tollerud, Geha, Grcevich, Putman  \&
  Stern}{Tollerud et~al.}{2014}]{tollerud2014}
Tollerud E.~J.,  Geha M.~C.,  Grcevich J.,  Putman M.~E.,   Stern D.,  2014,
  The Astrophysical Journal Letters, 798, L21

\bibitem[\protect\citeauthoryear{{Torrealba} et~al.,}{{Torrealba}
  et~al.}{2016}]{Torrealba16a}
{Torrealba} G.,  et~al., 2016, \mn@doi [\mnras] {10.1093/mnras/stw2051}, \href
  {https://ui.adsabs.harvard.edu/abs/2016MNRAS.463..712T} {463, 712}

\bibitem[\protect\citeauthoryear{{Torrealba} et~al.,}{{Torrealba}
  et~al.}{2018a}]{Torrealba2018b}
{Torrealba} G.,  et~al., 2018a, arXiv e-prints, \href
  {http://adsabs.harvard.edu/abs/2018arXiv181104082T} {}

\bibitem[\protect\citeauthoryear{{Torrealba} et~al.,}{{Torrealba}
  et~al.}{2018b}]{Torrealba2018a}
{Torrealba} G.,  et~al., 2018b, \mn@doi [\mnras] {10.1093/mnras/sty170}, \href
  {https://ui.adsabs.harvard.edu/abs/2018MNRAS.475.5085T} {475, 5085}

\bibitem[\protect\citeauthoryear{Van~Dokkum, Abraham, Merritt, Zhang, Geha  \&
  Conroy}{Van~Dokkum et~al.}{2015}]{van2015forty}
Van~Dokkum P.~G.,  Abraham R.,  Merritt A.,  Zhang J.,  Geha M.,   Conroy C.,
  2015, The Astrophysical Journal Letters, 798, L45

\bibitem[\protect\citeauthoryear{Viel, Becker, Bolton  \& Haehnelt}{Viel
  et~al.}{2013}]{Viel_2013}
Viel M.,  Becker G.~D.,  Bolton J.~S.,   Haehnelt M.~G.,  2013, \mn@doi
  [Physical Review D] {10.1103/physrevd.88.043502}, 88

\bibitem[\protect\citeauthoryear{{Vogelsberger} et~al.,}{{Vogelsberger}
  et~al.}{2014}]{Vogelsberger2014}
{Vogelsberger} M.,  et~al., 2014, \mn@doi [\mnras] {10.1093/mnras/stu1536},
  \href {https://ui.adsabs.harvard.edu/abs/2014MNRAS.444.1518V} {444, 1518}

\bibitem[\protect\citeauthoryear{{Vogelsberger}, {Zavala}, {Cyr-Racine},
  {Pfrommer}, {Bringmann}  \& {Sigurdson}}{{Vogelsberger}
  et~al.}{2016}]{Vogelsberger2016}
{Vogelsberger} M.,  {Zavala} J.,  {Cyr-Racine} F.-Y.,  {Pfrommer} C.,
  {Bringmann} T.,   {Sigurdson} K.,  2016, \mn@doi [\mnras]
  {10.1093/mnras/stw1076}, \href
  {http://adsabs.harvard.edu/abs/2016MNRAS.460.1399V} {460, 1399}

\bibitem[\protect\citeauthoryear{{Walsh}, {Willman}  \& {Jerjen}}{{Walsh}
  et~al.}{2009}]{Walsh2009}
{Walsh} S.~M.,  {Willman} B.,   {Jerjen} H.,  2009, \mn@doi [\aj]
  {10.1088/0004-6256/137/1/450}, \href
  {http://adsabs.harvard.edu/abs/2009AJ....137..450W} {137, 450}

\bibitem[\protect\citeauthoryear{Walter, Brinks, De~Blok, Bigiel, Kennicutt~Jr,
  Thornley  \& Leroy}{Walter et~al.}{2008}]{walter2008things}
Walter F.,  Brinks E.,  De~Blok W.,  Bigiel F.,  Kennicutt~Jr R.~C.,  Thornley
  M.~D.,   Leroy A.,  2008, The Astronomical Journal, 136, 2563

\bibitem[\protect\citeauthoryear{{Wechsler} \& {Tinker}}{{Wechsler} \&
  {Tinker}}{2018}]{WechslerTinker2018}
{Wechsler} R.~H.,  {Tinker} J.~L.,  2018, \mn@doi [Annual Review of Astronomy
  and Astrophysics] {10.1146/annurev-astro-081817-051756}, \href
  {https://ui.adsabs.harvard.edu/\#abs/2018ARA&A..56..435W} {56, 435}

\bibitem[\protect\citeauthoryear{{Weinberg}, {Hernquist}  \& {Katz}}{{Weinberg}
  et~al.}{1997}]{weinberg1997}
{Weinberg} D.~H.,  {Hernquist} L.,   {Katz} N.,  1997, \mn@doi [\apj]
  {10.1086/303683}, \href
  {https://ui.adsabs.harvard.edu/abs/1997ApJ...477....8W} {477, 8}

\bibitem[\protect\citeauthoryear{Weinberg, Bullock, Governato, de Naray  \&
  Peter}{Weinberg et~al.}{2015}]{Weinberg2015}
Weinberg D.~H.,  Bullock J.~S.,  Governato F.,  de Naray R.~K.,   Peter A.~H.,
  2015, Proceedings of the National Academy of Sciences, 112, 12249

\bibitem[\protect\citeauthoryear{Weisz et~al.,}{Weisz et~al.}{2011}]{weisz2011}
Weisz D.~R.,  et~al., 2011, The Astrophysical Journal, 743, 8

\bibitem[\protect\citeauthoryear{{Weisz}, {Dolphin}, {Skillman}, {Holtzman},
  {Gilbert}, {Dalcanton}  \& {Williams}}{{Weisz} et~al.}{2014}]{weisz2014b}
{Weisz} D.~R.,  {Dolphin} A.~E.,  {Skillman} E.~D.,  {Holtzman} J.,  {Gilbert}
  K.~M.,  {Dalcanton} J.~J.,   {Williams} B.~F.,  2014, \mn@doi [\apj]
  {10.1088/0004-637X/789/2/147}, \href
  {http://adsabs.harvard.edu/abs/2014ApJ...789..147W} {789, 147}

\bibitem[\protect\citeauthoryear{{Wetzel}, {Tollerud}  \& {Weisz}}{{Wetzel}
  et~al.}{2015}]{wetzel2015}
{Wetzel} A.~R.,  {Tollerud} E.~J.,   {Weisz} D.~R.,  2015, \mn@doi [\apjl]
  {10.1088/2041-8205/808/1/L27}, \href
  {http://adsabs.harvard.edu/abs/2015ApJ...808L..27W} {808, L27}

\bibitem[\protect\citeauthoryear{Wetzel, Hopkins, Kim, Faucher-Giguere, Keres
  \& Quataert}{Wetzel et~al.}{2016}]{Wetzel:2016wro}
Wetzel A.~R.,  Hopkins P.~F.,  Kim J.-h.,  Faucher-Giguere C.-A.,  Keres D.,
  Quataert E.,  2016, \mn@doi [\apj] {10.3847/2041-8205/827/2/L23}, 827, L23

\bibitem[\protect\citeauthoryear{{Wheeler} et~al.,}{{Wheeler}
  et~al.}{2019}]{Wheeler2019}
{Wheeler} C.,  et~al., 2019, \mn@doi [\mnras] {10.1093/mnras/stz2887}, \href
  {https://ui.adsabs.harvard.edu/abs/2019MNRAS.490.4447W} {490, 4447}

\bibitem[\protect\citeauthoryear{{Willman} et~al.,}{{Willman}
  et~al.}{2005}]{Willman05a}
{Willman} B.,  et~al., 2005, \mn@doi [\aj] {10.1086/430214}, \href
  {http://adsabs.harvard.edu/abs/2005AJ....129.2692W} {129, 2692}

\bibitem[\protect\citeauthoryear{{Wright} et~al.,}{{Wright}
  et~al.}{2010}]{WISE2010A}
{Wright} E.~L.,  et~al., 2010, \mn@doi [\aj] {10.1088/0004-6256/140/6/1868},
  \href {https://ui.adsabs.harvard.edu/abs/2010AJ....140.1868W} {140, 1868}

\bibitem[\protect\citeauthoryear{{York} et~al.,}{{York}
  et~al.}{2000}]{york2000}
{York} D.~G.,  et~al., 2000, \mn@doi [\aj] {10.1086/301513}, \href
  {http://adsabs.harvard.edu/abs/2000AJ....120.1579Y} {120, 1579}

\bibitem[\protect\citeauthoryear{{Zu} \& {Mandelbaum}}{{Zu} \&
  {Mandelbaum}}{2015}]{Zu2015}
{Zu} Y.,  {Mandelbaum} R.,  2015, \mn@doi [\mnras] {10.1093/mnras/stv2062},
  \href {http://adsabs.harvard.edu/abs/2015MNRAS.454.1161Z} {454, 1161}

\bibitem[\protect\citeauthoryear{{Zucker} et~al.,}{{Zucker}
  et~al.}{2004}]{Zucker04}
{Zucker} D.~B.,  et~al., 2004, \mn@doi [\apjl] {10.1086/424691}, \href
  {http://adsabs.harvard.edu/abs/2004ApJ...612L.121Z} {612, L121}

\bibitem[\protect\citeauthoryear{van~der Burg, Muzzin  \& Hoekstra}{van~der
  Burg et~al.}{2016}]{van_der_Burg_2016}
van~der Burg R. F.~J.,  Muzzin A.,   Hoekstra H.,  2016, \mn@doi [Astronomy &
  Astrophysics] {10.1051/0004-6361/201628222}, 590, A20

\makeatother
\end{thebibliography}


\vfill\null
\columnbreak
\appendix

\begin{table}
\caption{Completeness results after correcting for the area lost due to masking}
\centering 
\begin{tabular}{|r c r S S|}
\hline\hline
\multicolumn{1}{c}{} $\text{M}_{\text{V}}$ & $\mu_{\text{V}}$  & $\text{age}$ & $\text{metallicity}$  & $\text{completeness}$ \\
 & {\centering(\surfb{})}  &  & {\centering($\text{Z}/\text{Z}_{\odot}$)} & {(\%)}\\\hline
$-9$ & 25 & 100M & 0.01 & 35 \\ 
$-9$ & 25 & 100M & 0.1 & 35 \\ 
$-9$ & 25 & 1G & 0.01 & 35 \\ 
$-9$ & 25 & 1G & 0.1 & 35 \\ 
$-9$ & 25 & 10G & 0.01 & 63 \\ 
$-9$ & 25 & 10G & 0.1 & 63 \\ 
$-9$ & 26 & 100M & 0.01 & 34 \\ 
$-9$ & 26 & 100M & 0.1 & 34 \\ 
$-9$ & 26 & 1G & 0.01 & 34 \\ 
$-9$ & 26 & 1G & 0.1 & 33 \\ 
$-9$ & 26 & 10G & 0.01 & 58 \\ 
$-9$ & 26 & 10G & 0.1 & 58 \\ 
$-9$ & 27 & 100M & 0.01 & 22 \\ 
$-9$ & 27 & 100M & 0.1 & 26 \\ 
$-9$ & 27 & 1G & 0.01 & 26 \\ 
$-9$ & 27 & 1G & 0.1 & 24 \\ 
$-9$ & 27 & 10G & 0.01 & 48 \\ 
$-9$ & 27 & 10G & 0.1 & 48 \\ 
$-9$ & 28 & 100M & 0.01 & 6 \\ 
$-9$ & 28 & 100M & 0.1 & 9 \\ 
$-9$ & 28 & 1G & 0.01 & 10 \\ 
$-9$ & 28 & 1G & 0.1 & 10 \\ 
$-9$ & 28 & 10G & 0.01 & 17 \\ 
$-9$ & 28 & 10G & 0.1 & 20 \\ 
$-8$ & 26 & 100M & 0.01 & 35 \\ 
$-8$ & 26 & 100M & 0.1 & 32 \\ 
$-8$ & 26 & 1G & 0.01 & 34 \\ 
$-8$ & 26 & 1G & 0.1 & 34 \\ 
$-8$ & 26 & 10G & 0.01 & 60 \\ 
$-8$ & 26 & 10G & 0.1 & 60 \\ 
$-8$ & 27 & 100M & 0.01 & 30 \\ 
$-8$ & 27 & 100M & 0.1 & 25 \\ 
$-8$ & 27 & 1G & 0.01 & 27 \\ 
$-8$ & 27 & 1G & 0.1 & 28 \\ 
$-8$ & 27 & 10G & 0.01 & 46 \\ 
$-8$ & 27 & 10G & 0.1 & 50 \\ 
$-8$ & 28 & 100M & 0.01 & 16 \\ 
$-8$ & 28 & 100M & 0.1 & 10 \\ 
$-8$ & 28 & 1G & 0.01 & 15 \\ 
$-8$ & 28 & 1G & 0.1 & 12 \\ 
$-8$ & 28 & 10G & 0.01 & 21 \\ 
$-8$ & 28 & 10G & 0.1 & 25 \\ 
$-7$ & 26 & 10G & 0.01 & 56 \\ 
$-7$ & 26 & 10G & 0.1 & 62 \\ 
$-7$ & 27 & 10G & 0.01 & 47 \\ 
$-7$ & 27 & 10G & 0.1 & 49 \\ 
$-7$ & 28 & 10G & 0.01 & 18 \\ 
$-7$ & 28 & 10G & 0.1 & 19 \\ 
$-6$ & 27 & 10G & 0.01 & 25 \\ 
$-6$ & 27 & 10G & 0.1 & 26 \\ 
$-6$ & 28 & 10G & 0.01 & 22 \\ 
$-6$ & 28 & 10G & 0.1 & 24 \\ 
\hline
\end{tabular}
\end{table}

\begin{table}
\caption{Completeness results in the unmasked regions.}
\centering 
\begin{tabular}{|r c r S S|}
\hline\hline
\multicolumn{1}{c}{} $\text{M}_{\text{V}}$ & $\mu_{\text{V}}$  & $\text{age}$ &$\text{metallicity}$  & $\text{completeness}$ \\
 & {\centering(\surfb{})}  &  & {\centering($\text{Z}/\text{Z}_{\odot}$)} & {(\%)}\\
\hline
$-9$ & 25 & 100M & 0.01 & 88 \\ 
$-9$ & 25 & 100M & 0.1 & 88 \\ 
$-9$ & 25 & 1G & 0.01 & 88 \\ 
$-9$ & 25 & 1G & 0.1 & 87 \\ 
$-9$ & 25 & 10G & 0.01 & 88 \\ 
$-9$ & 25 & 10G & 0.1 & 87 \\ 
$-9$ & 26 & 100M & 0.01 & 85 \\ 
$-9$ & 26 & 100M & 0.1 & 83 \\ 
$-9$ & 26 & 1G & 0.01 & 84 \\ 
$-9$ & 26 & 1G & 0.1 & 83 \\ 
$-9$ & 26 & 10G & 0.01 & 81 \\ 
$-9$ & 26 & 10G & 0.1 & 82 \\ 
$-9$ & 27 & 100M & 0.01 & 54 \\ 
$-9$ & 27 & 100M & 0.1 & 66 \\ 
$-9$ & 27 & 1G & 0.01 & 67 \\ 
$-9$ & 27 & 1G & 0.1 & 62 \\ 
$-9$ & 27 & 10G & 0.01 & 66 \\ 
$-9$ & 27 & 10G & 0.1 & 66 \\ 
$-9$ & 28 & 100M & 0.01 & 15 \\ 
$-9$ & 28 & 100M & 0.1 & 22 \\ 
$-9$ & 28 & 1G & 0.01 & 27 \\ 
$-9$ & 28 & 1G & 0.1 & 25 \\ 
$-9$ & 28 & 10G & 0.01 & 23 \\ 
$-9$ & 28 & 10G & 0.1 & 28 \\ 
$-8$ & 26 & 100M & 0.01 & 88 \\ 
$-8$ & 26 & 100M & 0.1 & 80 \\ 
$-8$ & 26 & 1G & 0.01 & 85 \\ 
$-8$ & 26 & 1G & 0.1 & 84 \\ 
$-8$ & 26 & 10G & 0.01 & 84 \\ 
$-8$ & 26 & 10G & 0.1 & 84 \\ 
$-8$ & 27 & 100M & 0.01 & 75 \\ 
$-8$ & 27 & 100M & 0.1 & 65 \\ 
$-8$ & 27 & 1G & 0.01 & 71 \\ 
$-8$ & 27 & 1G & 0.1 & 70 \\ 
$-8$ & 27 & 10G & 0.01 & 62 \\ 
$-8$ & 27 & 10G & 0.1 & 69 \\ 
$-8$ & 28 & 100M & 0.01 & 42 \\ 
$-8$ & 28 & 100M & 0.1 & 25 \\ 
$-8$ & 28 & 1G & 0.01 & 39 \\ 
$-8$ & 28 & 1G & 0.1 & 30 \\ 
$-8$ & 28 & 10G & 0.01 & 28 \\ 
$-8$ & 28 & 10G & 0.1 & 33 \\ 
$-7$ & 26 & 10G & 0.01 & 78 \\ 
$-7$ & 26 & 10G & 0.1 & 86 \\ 
$-7$ & 27 & 10G & 0.01 & 66 \\ 
$-7$ & 27 & 10G & 0.1 & 68 \\ 
$-7$ & 28 & 10G & 0.01 & 25 \\ 
$-7$ & 28 & 10G & 0.1 & 27 \\ 
$-6$ & 27 & 10G & 0.01 & 35 \\ 
$-6$ & 27 & 10G & 0.1 & 38 \\ 
$-6$ & 28 & 10G & 0.01 & 30 \\ 
$-6$ & 28 & 10G & 0.1 & 33 \\
\hline
\end{tabular}
\end{table}

\clearpage

\begin{table}
\caption{{\tt{SExtractor}} parameters. These parameters are for chip 1. The detection parameters varied slightly from chip to chip. }
\centering 
\begin{tabular}{|c c r S S c|}
\hline\hline
\multicolumn{1}{c}{} $\text{M}_{\text{V}}$ & $\mu_{\text{V}}$  & age & $\text{metallicity}$  & $\text{DETECT}\_\text{MINAREA}$ & $\text{DETECT}\_\text{THRESH}$ \\
 & {\centering(\surfb{})}  &  & {\centering($\text{Z}/\text{Z}_{\odot}$)} & &\\[0.5ex]\hline
$-9$ & 25 & 100M & 0.01 & 500 & 4.0 \\ 
$-9$ & 25 & 100M & 0.1 & 400 & 4.0 \\ 
$-9$ & 25 & 1G & 0.01 & 400 & 4.0 \\ 
$-9$ & 25 & 1G & 0.1 & 450 & 3.0 \\ 
$-9$ & 25 & 10G & 0.01 & 500 & 2.0 \\ 
$-9$ & 25 & 10G & 0.1 & 500 & 5.0 \\ 
$-9$ & 26 & 100M & 0.01 & 800 & 1.0 \\ 
$-9$ & 26 & 100M & 0.1 & 700 & 2.0 \\ 
$-9$ & 26 & 1G & 0.01 & 700 & 2.0 \\ 
$-9$ & 26 & 1G & 0.1 & 600 & 2.0 \\ 
$-9$ & 26 & 10G & 0.01 & 800 & 3.0 \\ 
$-9$ & 26 & 10G & 0.1 & 800 & 4.0 \\ 
$-9$ & 27 & 100M & 0.01 & 1200 & 0.5 \\ 
$-9$ & 27 & 100M & 0.1 & 1200 & 0.5 \\ 
$-9$ & 27 & 1G & 0.01 & 1200 & 1.0 \\ 
$-9$ & 27 & 1G & 0.1 & 1200 & 1.0 \\ 
$-9$ & 27 & 10G & 0.01 & 1200 & 2.0 \\ 
$-9$ & 27 & 10G & 0.1 & 1200 & 2.0 \\ 
$-9$ & 28 & 100M & 0.01 & 700 & 0.5 \\ 
$-9$ & 28 & 100M & 0.1 & 1100 & 0.5 \\ 
$-9$ & 28 & 1G & 0.01 & 1600 & 0.5 \\ 
$-9$ & 28 & 1G & 0.1 & 1700 & 0.5 \\ 
$-9$ & 28 & 10G & 0.01 & 1700 & 0.5 \\ 
$-9$ & 28 & 10G & 0.1 & 2000 & 0.5 \\ 
$-8$ & 26 & 100M & 0.01 & 400 & 2.0 \\ 
$-8$ & 26 & 100M & 0.1 & 500 & 1.0 \\ 
$-8$ & 26 & 1G & 0.01 & 500 & 1.0 \\ 
$-8$ & 26 & 1G & 0.1 & 500 & 1.0 \\ 
$-8$ & 26 & 10G & 0.01 & 500 & 2.0 \\ 
$-8$ & 26 & 10G & 0.1 & 500 & 2.0 \\ 
$-8$ & 27 & 100M & 0.01 & 700 & 0.5 \\ 
$-8$ & 27 & 100M & 0.1 & 800 & 0.5 \\ 
$-8$ & 27 & 1G & 0.01 & 800 & 0.5 \\ 
$-8$ & 27 & 1G & 0.1 & 800 & 1.0 \\ 
$-8$ & 27 & 10G & 0.01 & 600 & 1.0 \\ 
$-8$ & 27 & 10G & 0.1 & 500 & 2.0 \\ 
$-8$ & 28 & 100M & 0.01 & 1100 & 0.5 \\ 
$-8$ & 28 & 100M & 0.1 & 800 & 0.5 \\ 
$-8$ & 28 & 1G & 0.01 & 1000 & 0.5 \\ 
$-8$ & 28 & 1G & 0.1 & 700 & 0.5 \\ 
$-8$ & 28 & 10G & 0.01 & 900 & 0.5 \\ 
$-8$ & 28 & 10G & 0.1 & 800 & 1.0 \\ 
$-7$ & 26 & 10G & 0.01 & 250 & 2.0 \\ 
$-7$ & 26 & 10G & 0.1 & 200 & 3.0 \\ 
$-7$ & 27 & 10G & 0.01 & 250 & 1.0 \\ 
$-7$ & 27 & 10G & 0.1 & 350 & 1.0 \\ 
$-7$ & 28 & 10G & 0.01 & 400 & 0.5 \\ 
$-7$ & 28 & 10G & 0.1 & 400 & 0.5 \\ 
$-6$ & 27 & 10G & 0.01 & 200 & 1.0 \\ 
$-6$ & 27 & 10G & 0.1 & 200 & 1.0 \\ 
$-6$ & 28 & 10G & 0.01 & 100 & 1.0 \\ 
$-6$ & 28 & 10G & 0.1 & 100 & 1.0 \\
\hline

\end{tabular}
\end{table}

\label{lastpage}
\end{document}